\documentclass[journal=jpclcd,manuscript=article]{achemso}
\author{Sourin Dey}
\author{Nicholas Miklaucic}
\author{Sadman Sadeed Omee}
\author{Rongzhi Dong}
\author{Lai Wei}
\author{Qinyang Li}
\author{Nihang Fu}
\affiliation{Department of Computer Science and Engineering, University of South Carolina, Columbia, SC, USA}
\author{Jianjun Hu}
\email{jianjunh@cse.sc.edu}
\affiliation{Department of Computer Science and Engineering, University of South Carolina, Columbia, SC, USA}

\title{Data-Driven Topological Analysis of Polymorphic Crystal Structures}

\usepackage{algorithm}
\usepackage{algorithmic}
\usepackage{amsmath}
\usepackage{graphicx}
\usepackage{subcaption}
\usepackage{caption}
\usepackage{booktabs}
\usepackage{multirow}
\usepackage{color}
\usepackage{colortbl}
\usepackage{placeins}
\usepackage{array}
\usepackage{soul}
\usepackage{hyperref}
\usepackage{xr}

\begin{document}

\begin{tocentry}
\vspace{0.75cm}
\includegraphics[width=5.1cm]{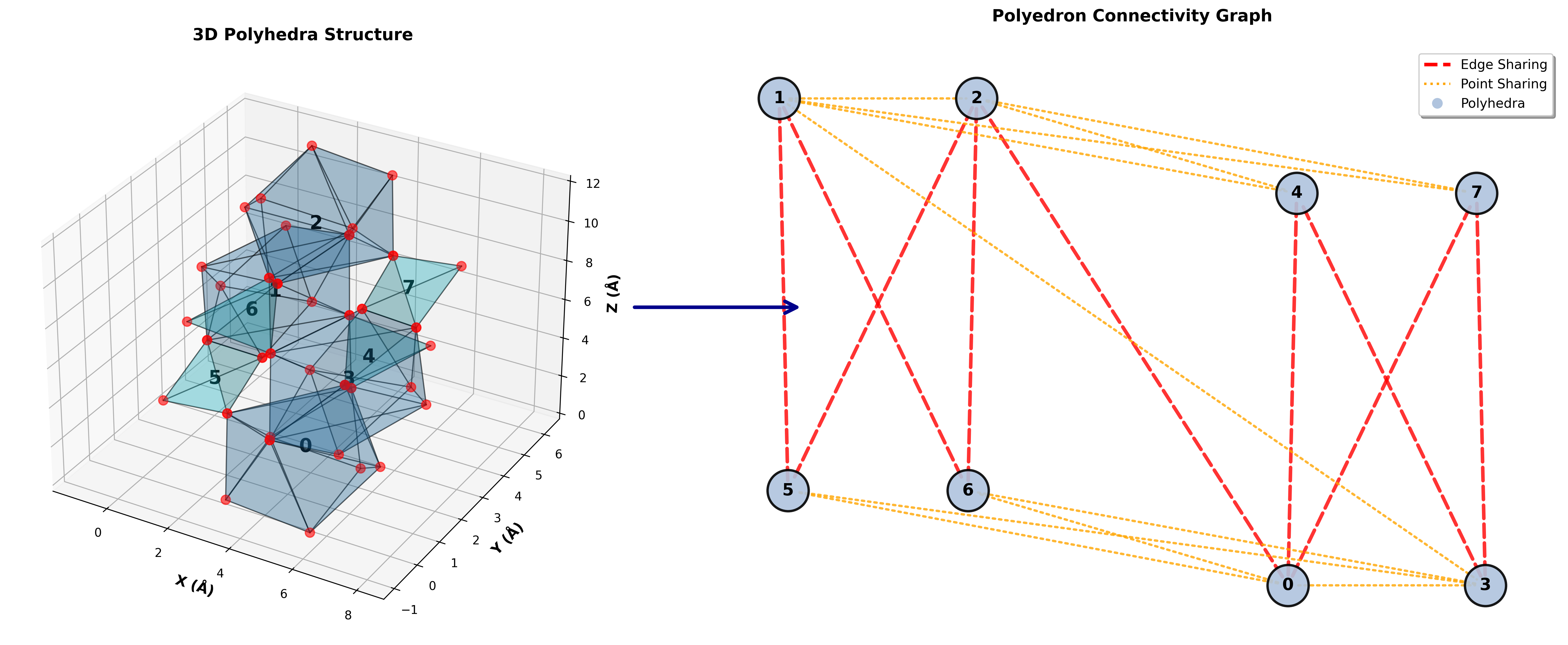}
\end{tocentry}

\begin{abstract}

Polymorphism, the ability of a compound to crystallize in multiple distinct structures, which plays a vital role in determining the physical, chemical, and functional properties of materials.
Accurate identification and prediction of polymorphic structures are critical for materials design, drug development, and device optimization, as unknown or overlooked polymorphs may lead to unexpected performance or stability issues.
Despite its significance, predicting polymorphism directly from a chemical composition remains a challenging problem due to the complex interplay between molecular conformations, crystal packing, and symmetry constraints.
In this study, we conduct a comprehensive data-driven analysis of polymorphic materials from the Materials Project database, uncovering key statistical patterns in their composition, space group distributions, and polyhedral building blocks. We discover that frequent polymorph pairs across space groups, such as (71, 225), display recurring topological motifs that persist across different compounds, highlighting topology — not symmetry alone — as a key factor in polymorphic recurrence. We reveal that many polymorphs exhibit consistent local polyhedral environments despite differences in their symmetry or packing. Additionally, by constructing polyhedron connectivity graphs and embedding their topology, we successfully cluster polymorphs and structurally similar materials even across different space groups, demonstrating that topological similarity serves as a powerful descriptor for polymorphic behavior.
Our findings provide new insights into the structural characteristics of polymorphic materials and demonstrate the potential of data mining and machine learning for accelerating polymorph discovery and design.
\end{abstract}

\section{Introduction}

Polymorphism in materials refers to the ability of a substance to exist in multiple crystalline forms, known as polymorphs, each with distinct atomic or molecular arrangements in the crystal lattice.
This phenomenon is of significant importance in various fields, including materials science, pharmaceuticals, and nanotechnology, as polymorphs can exhibit markedly different physical, chemical, and mechanical properties, such as solubility, stability, and electronic behavior.
For example, zinc oxide exhibits polymorphism with different crystal structures that can influence its optical and electronic properties, making it valuable in applications such as piezoelectric transducers, solar cells, and sensors \cite{djurivsic2006optical}.
Similarly, silicon, which is commonly used in electronics, exhibits different polymorphs that can impact its electronic and thermal properties, influencing the performance of semiconductors \cite{rapp2015experimental}.
In the pharmaceutical industry, polymorphism is crucial for optimizing drug formulations, as different polymorphs of a compound can influence its availability, efficacy, and stability \cite{brittain1999polymorphism}.
There are two primary types of polymorphism: enantiotropic polymorphism, where polymorphs can reversibly transform with changes in temperature or pressure, and monotropic polymorphism, where transformation occurs irreversibly under specific conditions.
In materials science and nanotechnology, controlling polymorphism enables the design of materials with tailored optical, electronic, and magnetic properties for applications in semiconductors, photovoltaics, and catalysts \cite{brog2013polymorphism,gentili2019polymorphism,anjaneyulu2024exploring,chung2016polymorphism}.
One such example is controlling In$_{2}$Se$_{3}$. Instead of applying a mechanically driven process, a recent study \cite{modi2024electrically} has found an energy-efficient approach by employing an electric field-induced conversion of In$_{2}$Se$_{3}$ into a novel ferroic $\beta$'' phase, which can potentially revolutionize the data storage process.
Different polymorph phases, such as the R-phase and T-phase of materials, are quite important for tuning the properties of actuators, sensors, and memory devices because they demonstrate distinct ferroelectric as well as piezoelectric properties in those different phases \cite{chen2016giant}.
Thus, fundamental insights into these structural transformations are essential for discovering and accessing useful states of materials.
The vast structural and polymorphic varieties found in chalcogenide semiconductors like the AMM\'Q4 systems further underscore this importance \cite{friedrich2021vast}.
The ability to predict polymorphic behavior is, therefore, crucial for the development of advanced materials with desired characteristics and for improving the efficiency and performance of a wide range of industrial applications.
Polymorphism also plays a vital role in crystal structure generation, a major topic in materials research \cite{zeni2023mattergen,lin2024equivariant,jiao2023crystal,dong2024generative,wei2024crystal}, as accounting for multiple possible forms enables a comprehensive exploration of a material's configurational landscape.
Incorporating polymorphic variants thus enhances the accuracy of structure prediction methods and aids in identifying energetically favorable or functionally superior phases. A recent study has emphasized the importance of considering polymorphism in the crystal structure prediction (CSP) problem by introducing a novel genetic algorithm specifically designed to efficiently explore polymorphic variants \cite{omee2025polymorphism}.
\\

Crystal polymorphism has been studied over the years due to its critical implications in materials science, pharmaceuticals, and solid-state  \cite{yu2010polymorphism,aulakh2015importance,mizuno2016control,omee2025polymorphism}.
Much of this research has traditionally focused on understanding how molecular features influence the propensity of a compound to exhibit polymorphism.
In particular, molecular flexibility and conformational diversity have been recognized as important factors contributing to the formation of different crystal structures.
Conformational effects on crystal polymorphism have been systematically studied, with a detailed analysis of how molecular flexibility, conformer populations, and intramolecular interactions influence polymorph formation presented by Cruz-Cabeza and Bernstein \cite{cruz2014conformational}.
Nevertheless, the relationship between molecular characteristics and the occurrence of crystal polymorphism remains complex.
Research suggests that molecular size or flexibility alone does not necessarily determine polymorphic behavior, as both large, flexible molecules and small, rigid ones can exhibit similar tendencies to form polymorphs.
Chiral molecules, however, are generally observed to be less prone to polymorphism compared to their non-chiral counterparts, and the presence of hydrogen-bonding functionalities slightly increases the likelihood of polymorphism \cite{cruz2015facts}.
Several case studies illustrate the compound-specific nature of polymorphism. For example, paracetamol is known to exhibit three polymorphs that differ in hydrogen bonding and crystal packing arrangements \cite{higashi2017recent}.
Carbamazepine displays at least four anhydrous polymorphs arising from conformational flexibility and multiple stable conformers in the solid state \cite{grzesiak2003comparison}.
The case of ritonavir further underscores the unpredictable and impactful nature of polymorphism: a previously unobserved, more stable polymorph (Form II) emerged after the drug’s commercialization, rendering the original formulation less soluble and leading to major manufacturing setbacks and product recalls \cite{bauer2001ritonavir}.
Beyond molecular considerations, recent studies have emphasized the importance of crystal-level classification in understanding polymorphism prevalence.
Classifying crystal structures into subclasses such as anhydrates, salts, hydrates, non-hydrated solvates, and cocrystals has revealed substantial differences in polymorphism frequency across these types—a fact that had not been widely recognized in the literature until recently \cite{kersten2018survey}.
These recent studies are complemented by advancements in computational methods. For instance, PCRL employs a multi-modal pre-training framework that integrates detailed crystal structure data with compositional descriptors, allowing the model to learn representations where compositional information is implicitly linked to the diverse structural arrangements possible for that composition \cite{lee2023compositional}.
\\

In this work, we go beyond traditional symmetry-based analysis by introducing a topology-driven framework for polymorph characterization, centered on polyhedral connectivity and local structural motifs. Leveraging this approach, we conduct a large-scale data analysis of polymorphic materials from the Materials Project database, uncovering statistical patterns and distributions across composition, crystal structure, symmetry, oxidation states, and local polyhedral environments.
To gain deeper insights, we further categorize the polymorph data into representative material prototypes such as rocksalt, fluorite, chalcopyrite, perovskite, pyrochlore, spinel, and scheelite, enabling a systematic comparison of their structural characteristics.
Building on these observations, we construct polyhedron connectivity graphs that encode the topological relationships between local coordination polyhedra in each structure.
By embedding these graphs into a vector space, we identify clusters of topologically similar materials, including polymorphs spanning different space groups.
This analysis reveals that polyhedral connectivity, rather than symmetry alone, can serve as a robust descriptor for capturing structural similarity and polymorphic recurrence.
Furthermore, in specific chemical families such as lithium manganese cobalt oxides, we observe that polymorphs can retain remarkably consistent local topological patterns even when composed of different types of polyhedra, suggesting deeper structural constraints governing polymorphic stability in certain systems.
This polyhedron-centric perspective aligns with emerging trends in crystal structure modeling. Recent work by Yokoyama et al.
\cite{yokoyama2024crystal} demonstrated the potential of space-filling polyhedra as generative building blocks for designing crystal structures, offering a topological approach to structure generation.
Similarly, Zhu et al. \cite{zhu2017efficient} developed a crystal structure prediction method that uses rigid polyhedral units to guide efficient configuration space exploration.
These studies reinforce the broader utility of polyhedra not only in structural analysis, as demonstrated here, but also in the predictive modeling of polymorphs and crystal generation.
\section{Dataset Overview}
\label{sec:headings}

\subsection{Dataset Construction}

We collected a total of 56,920 crystal structures from the Materials Project \cite{jain2013commentary,horton2025accelerated} database by selecting entries that contain fewer than 25 atoms per unit cell to ensure computational tractability and focus on well-defined, representative prototypes. Among these, we identified 6,287 unique reduced chemical formulas that exhibit polymorphism, defined as having more than one distinct crystal structure reported for the same formula. These 6,287 polymorphic formulas account for a total of 19,049 polymorphic structure entries, forming the basis of our subsequent statistical and topological analysis.
\begin{figure}[h!]
    \centering
    \begin{minipage}[c]{0.51\textwidth}
        \centering
        \vspace*{2.5em}  %
        \includegraphics[width=\linewidth]{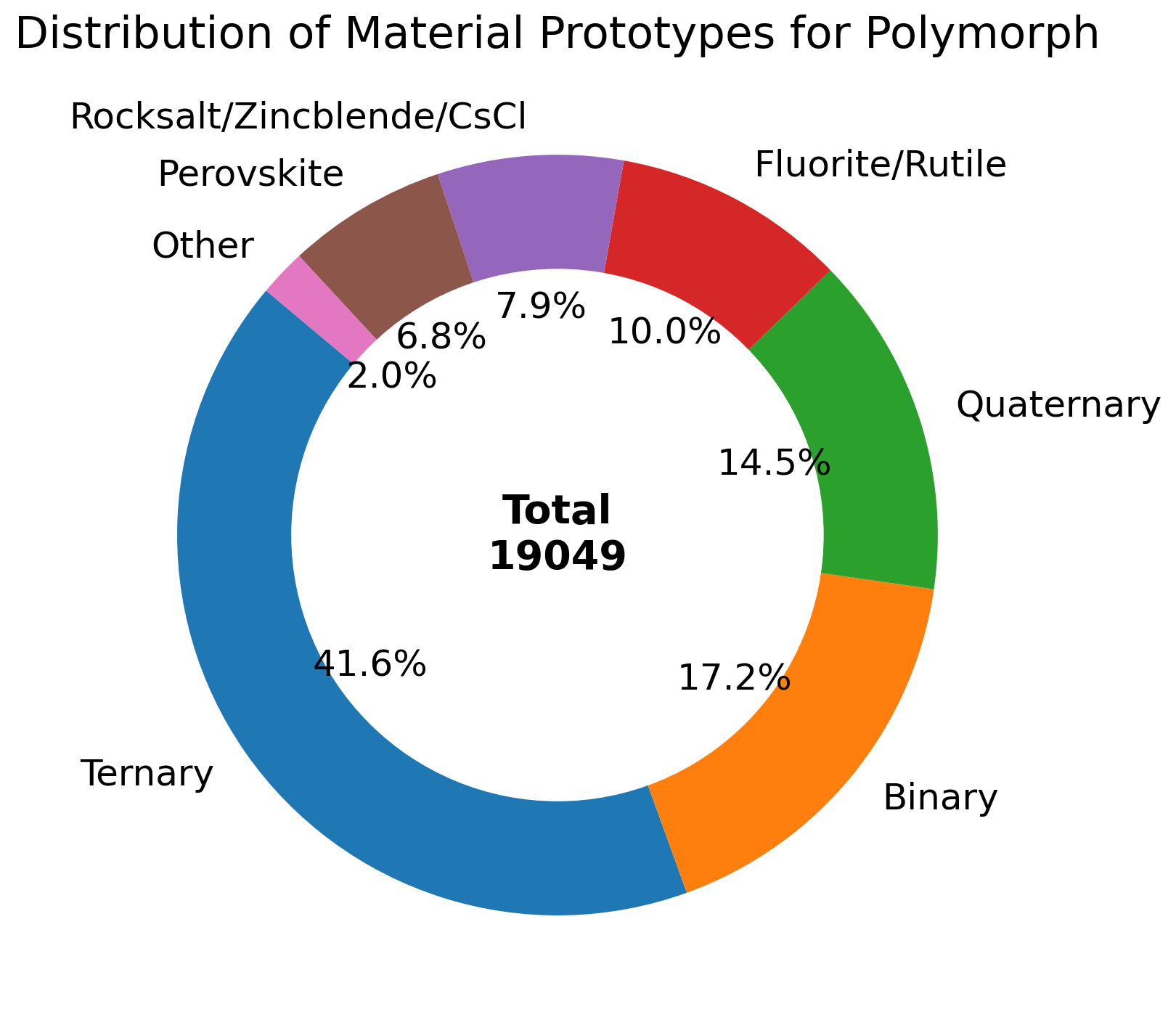}
        \caption*{(a)}
       
    \end{minipage}%
    \hfill
    \begin{minipage}[c]{0.45\textwidth}
        \centering
   
     \includegraphics[width=\linewidth]{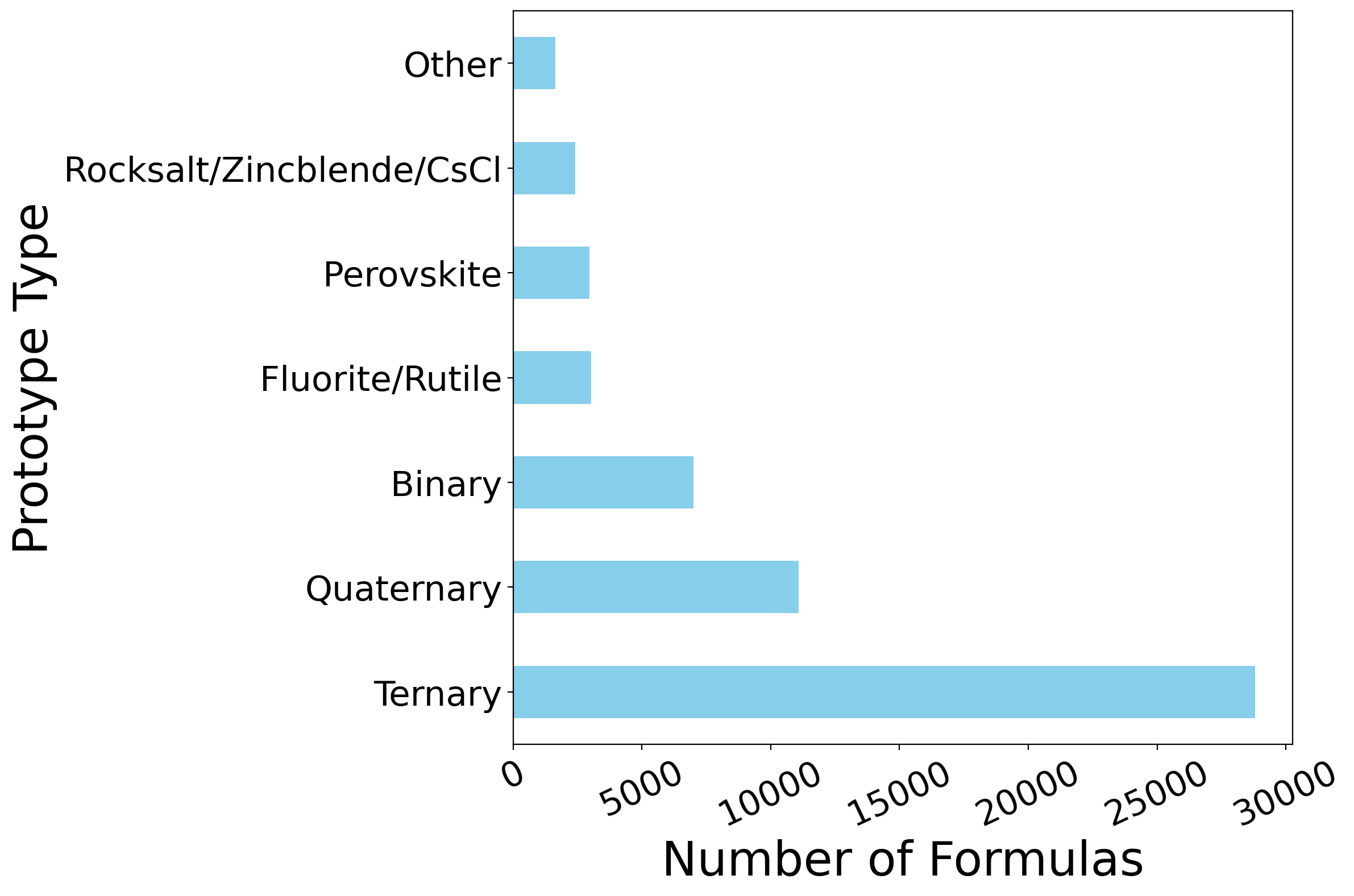}
        \caption*{(b)}
        \vspace{2em}
        
        \includegraphics[width=\linewidth]{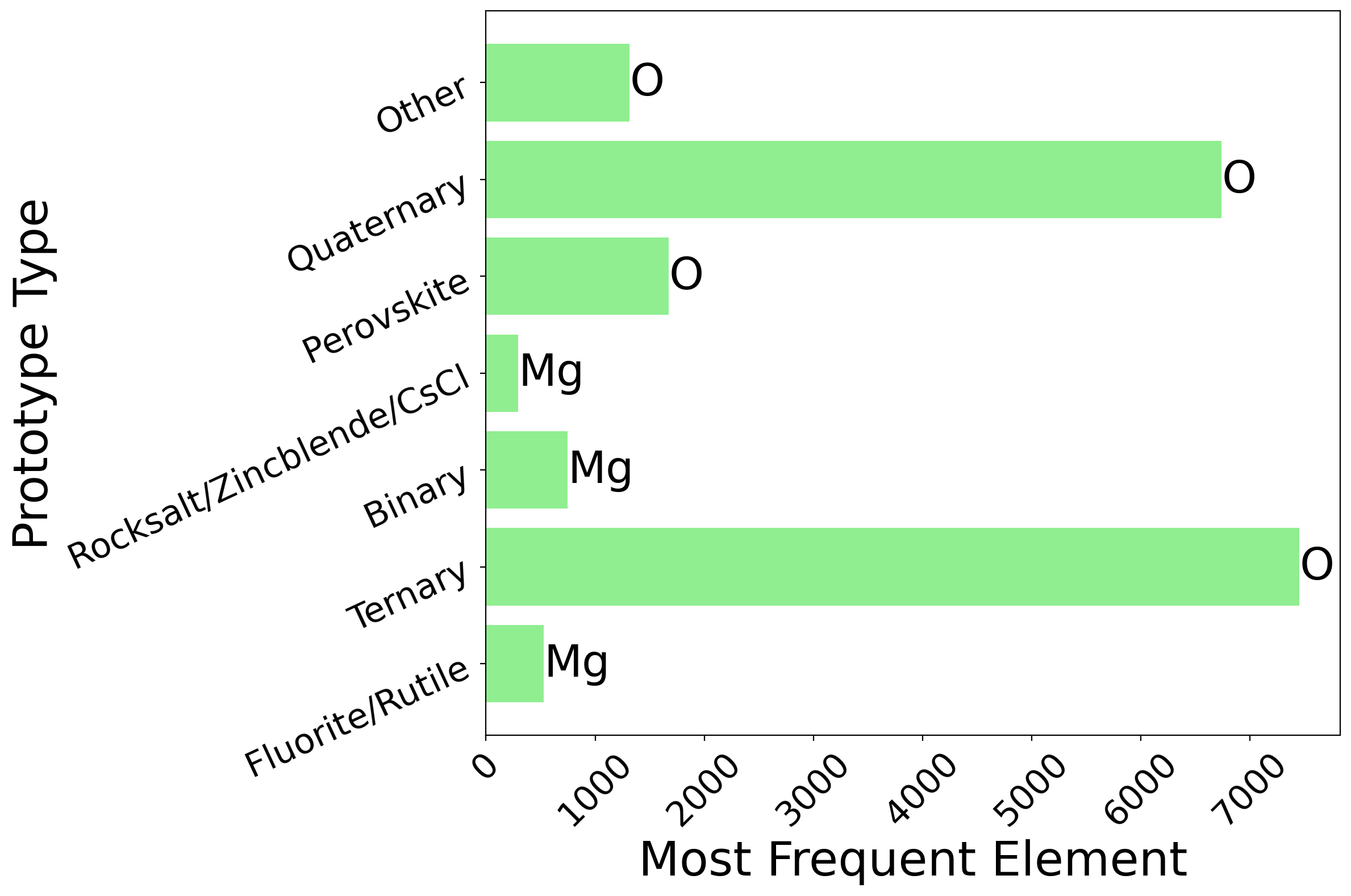}
        \caption*{(c)}
    \end{minipage}

    \caption{Distribution of material prototypes collected from the Materials Project database.
(a) Prototype categories in the dataset. (b) Prototype categories among only polymorphic structures.
(c) Most frequent element within each prototype category.}
     \label{fig:dataset_polymorph}
\end{figure}

\section{Results}

\subsection{Prototype-Specific Trends}

Figure~\ref{fig:dataset_polymorph} provides a breakdown of polymorphic entries by material prototype.
Figure~\ref{fig:dataset_polymorph}a shows the distribution of all collected structures by prototype class, with ternary compounds being the most common, followed by quaternary and binary materials.
\par
Figure~\ref{fig:dataset_polymorph}b focuses only on the polymorphic subset, revealing a similar trend where ternary systems dominate (41.6\%), followed by binary (17.2\%) and quaternary (14.5\%) entries.
Notably, distinct structural families such as Fluorite/Rutile and Perovskite also account for a significant share of polymorphs.
\par
Figure~\ref{fig:dataset_polymorph}c examines the most frequent elements found in each prototype.
Oxygen dominates in ternary, quaternary, and perovskite polymorphs, while magnesium is most common in binary, Fluorite/Rutile, and Rocksalt-type systems.
These trends suggest that oxygen-based coordination environments—particularly in metal oxides and fluorides—are key contributors to structural flexibility and the likelihood of polymorphic behavior.

\subsection{Composition and Symmetry Distributions}

\begin{figure}[th]
    \centering
    \begin{minipage}{0.49\textwidth}
        \centering
        \includegraphics[width=\textwidth]{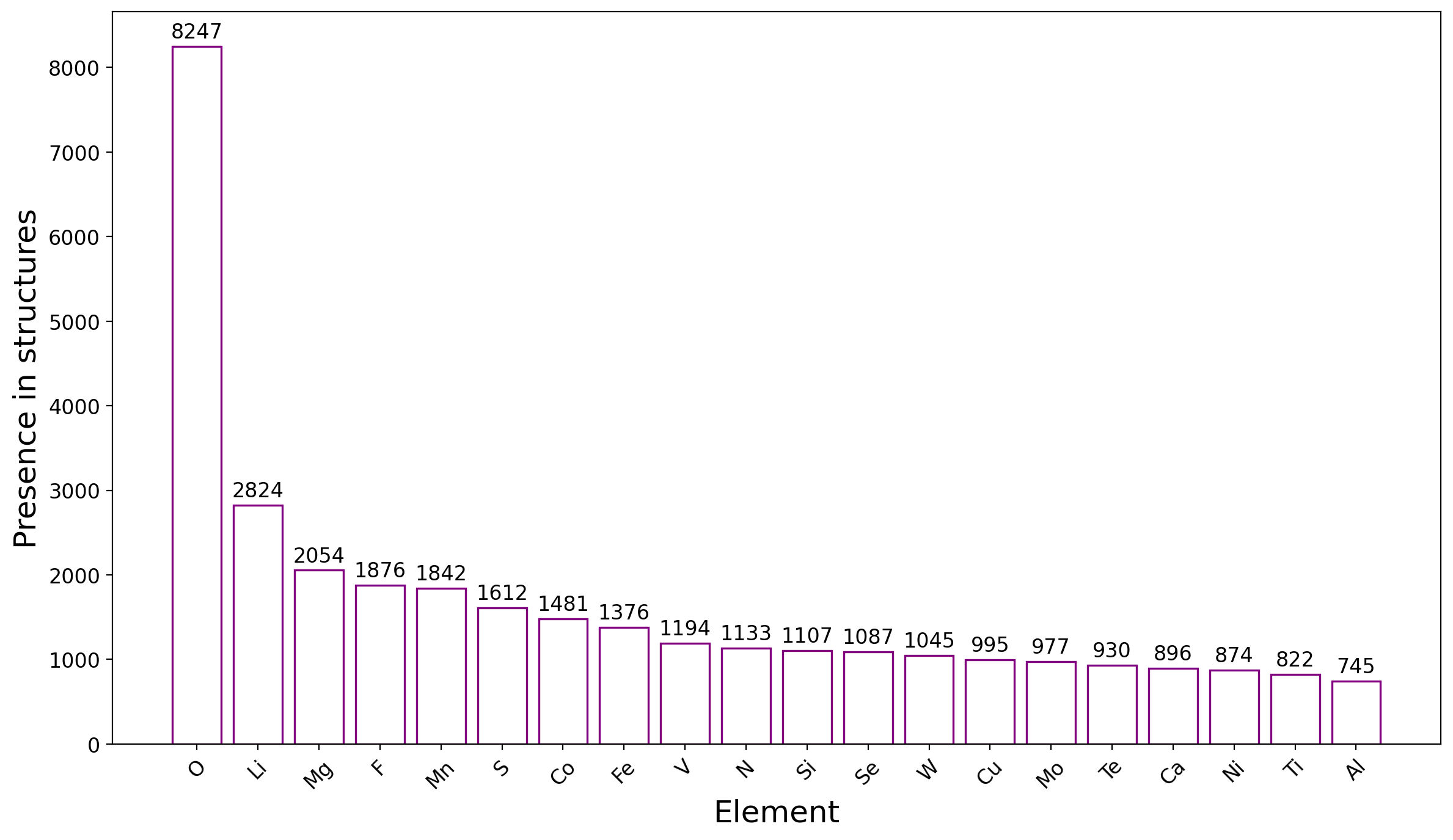}
        \subcaption[ \hspace{0.2 cm} a]{}
    \end{minipage}
    \begin{minipage}{0.49\textwidth}
        \centering
        \includegraphics[width=\textwidth]{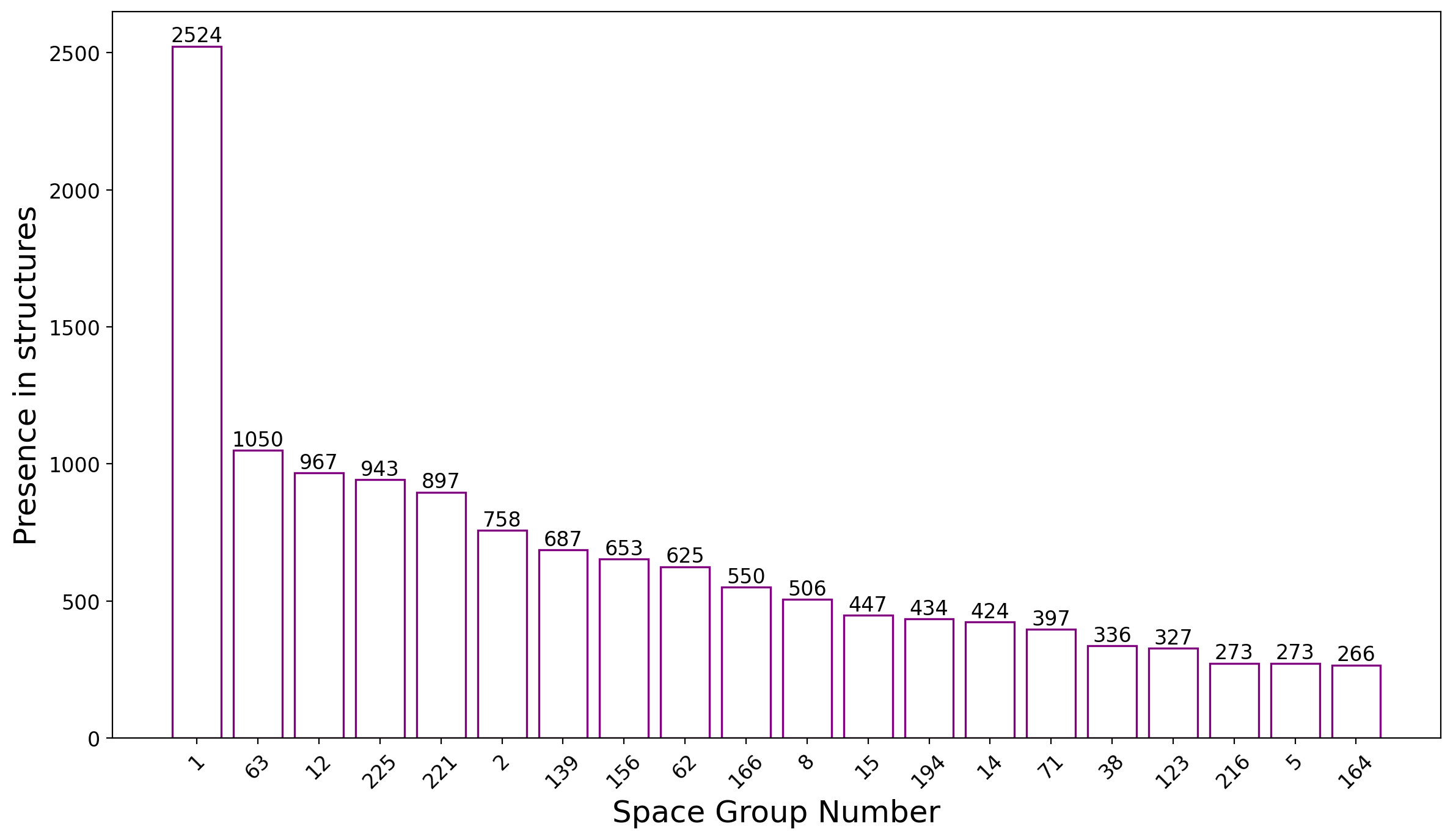}
    
    \subcaption[  b]{}
    \end{minipage}
    
    \vspace{0.3cm} %
    
    \begin{minipage}{0.37\textwidth} %
        \centering
        \includegraphics[width=\textwidth]{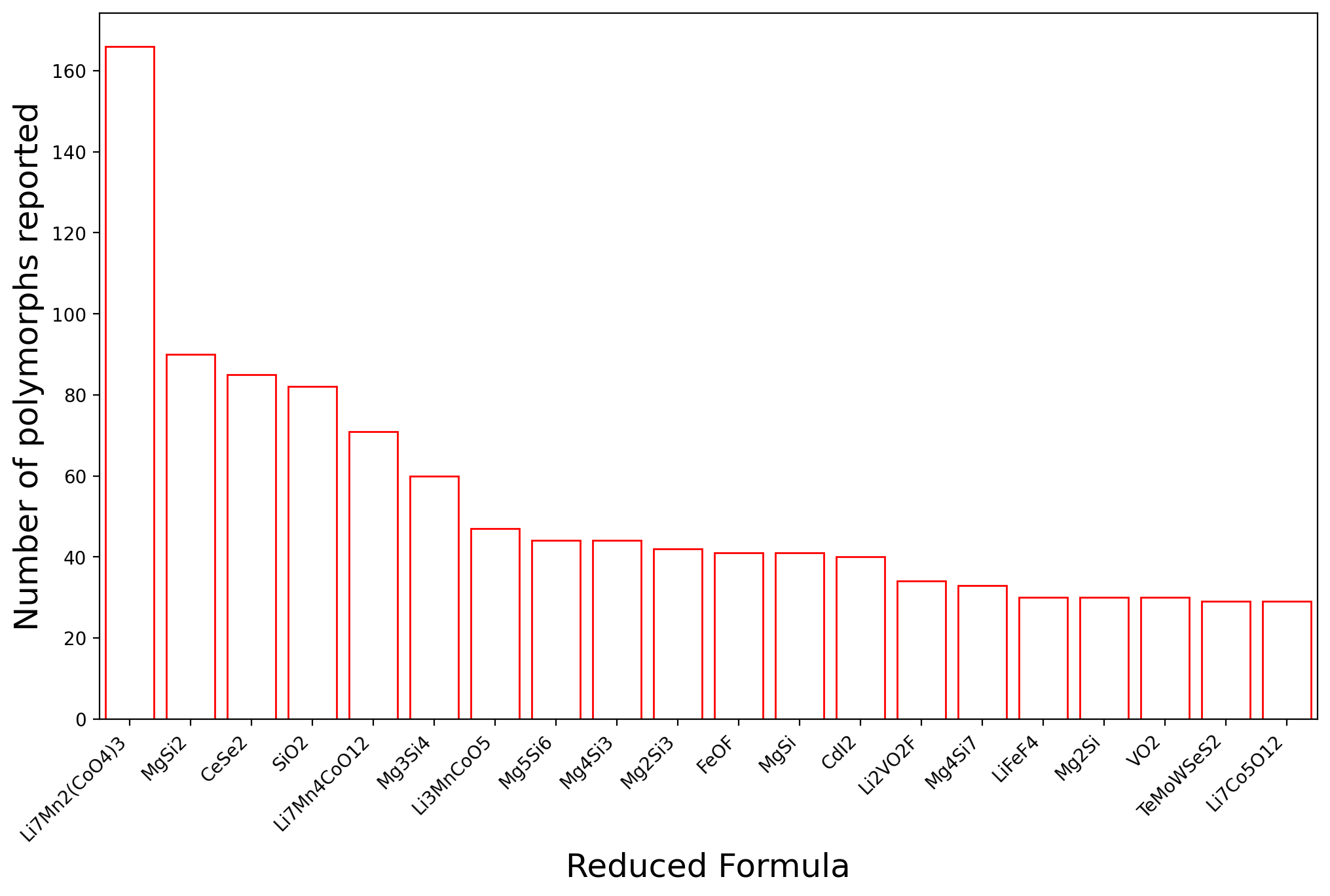}
        \subcaption[    c]{}
    \end{minipage}
    \begin{minipage}{0.25\textwidth}
        \centering
      
  \includegraphics[width=\textwidth]{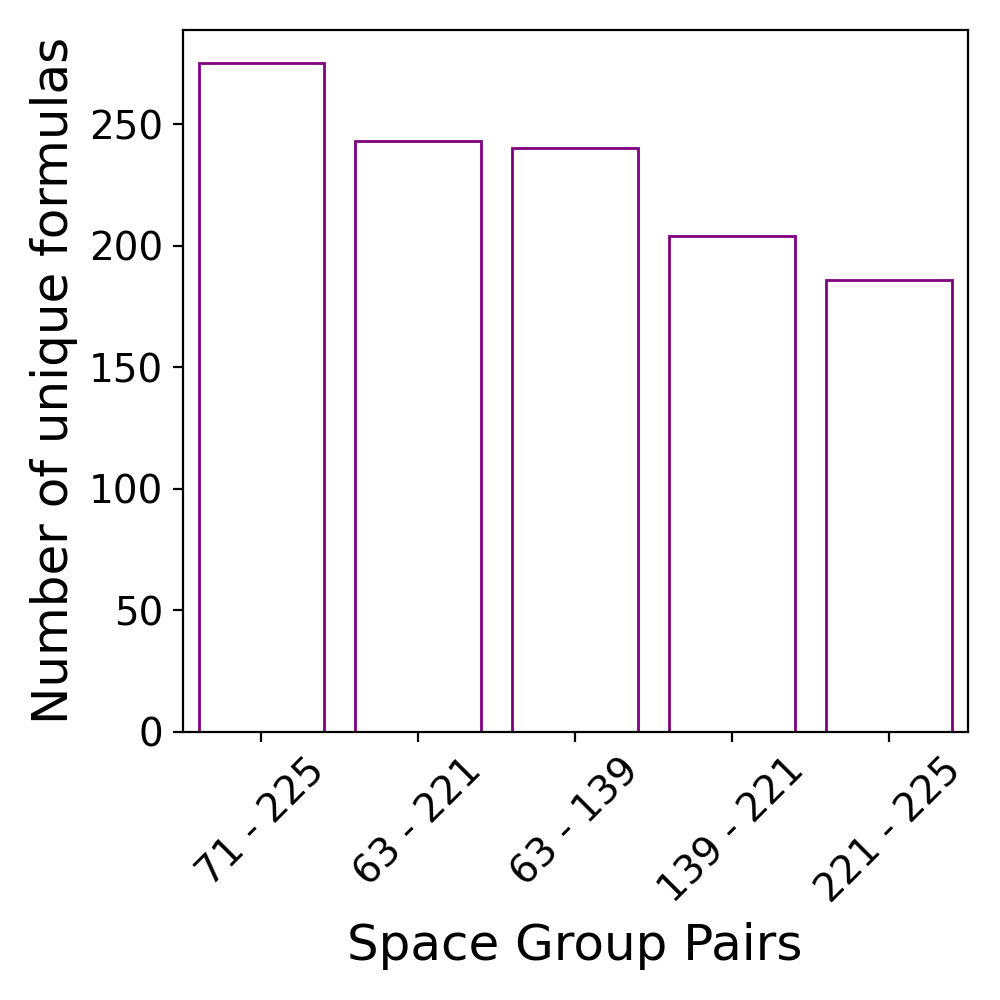}
        \subcaption[    d]{}
    \end{minipage}
    \begin{minipage}{0.32\textwidth}
        \centering
        \includegraphics[width=\textwidth]{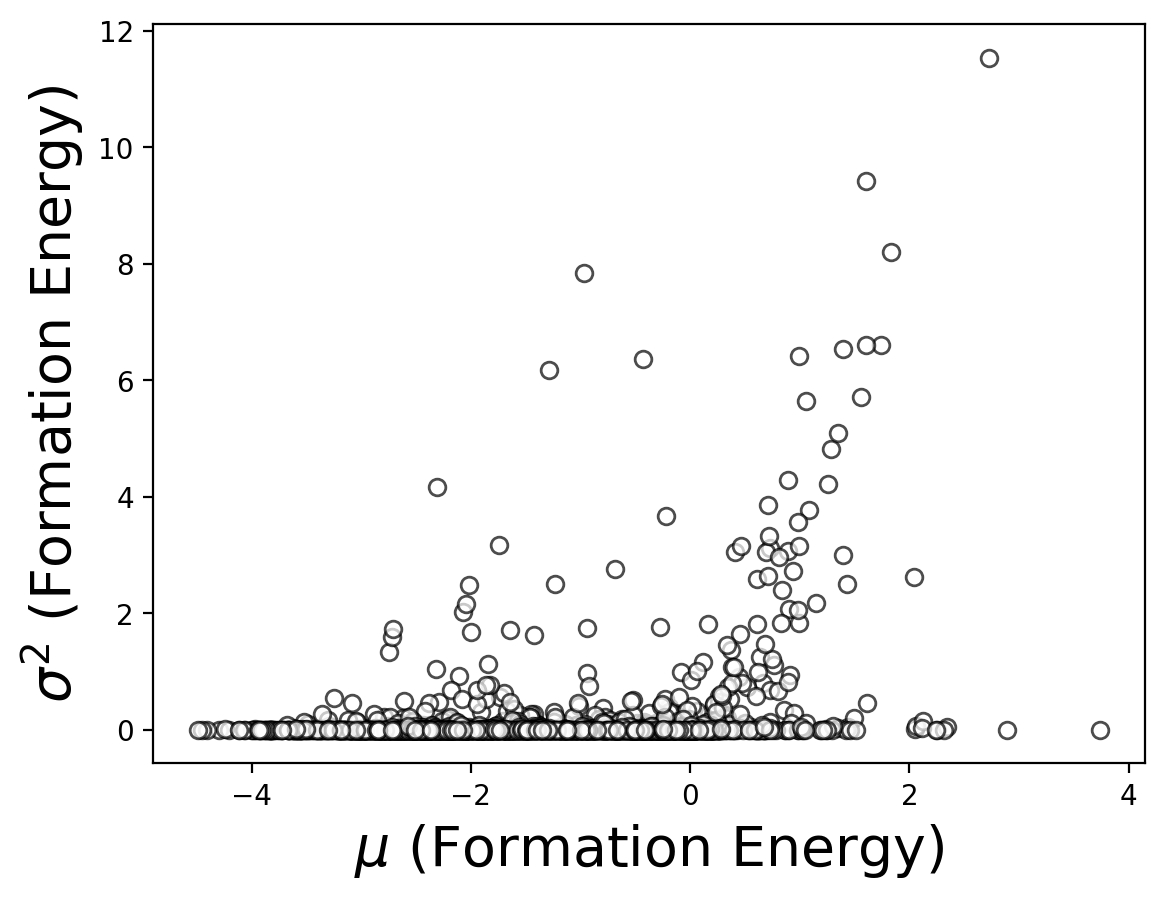}
        \subcaption[    e]{}
    \end{minipage}
    
    \caption{Exploration of polymorphs in the Materials Project dataset.
The analysis conducted on this dataset includes 6,287 polymorphs. The top-left figure shows the distribution of the most frequent elements in different polymorph structures, where Oxygen (O) is the most common (8,247 structures).
The top-right figure presents the top 20 most frequent space groups, with space group 1 (triclinic) being the most common (2,524 structures).
The bottom-left figure displays the top 20 formulas that have the maximum polymorph occurrences, and the bottom-right shows the top 5 most frequent space group pairs.
(e) Relationship between the mean and variance of formation energy among polymorphs.}
    \label{fig:combined_image}
\end{figure}

Figure~\ref{fig:combined_image}a shows the distribution of the most frequent elements found in polymorphic structures.
Oxygen (O) is by far the most prevalent, appearing in 8,247 structures, which suggests that oxide-based systems dominate the polymorph landscape.
This is followed by other common anions and small cations, indicating that ionic bonding environments frequently support multiple packing arrangements.
\par
In Figure~\ref{fig:combined_image}b, we show the 20 most frequent space groups among all polymorphic entries.
Space group 1 (triclinic, P1) appears most frequently, with 2,524 entries, followed by monoclinic and orthorhombic groups.
This distribution suggests that lower-symmetry structures are more likely to exhibit polymorphism, possibly due to increased configurational flexibility.
\par
Figure~\ref{fig:combined_image}c presents the top 20 reduced formulas with the highest number of reported polymorphs.
For instance, Li$_{7}$Mn$_{2}$(CoO$_{4}$)$_{3}$ appears in over 160 unique crystal configurations, showcasing the extent of structural richness within some composition families.
\par
Figure~\ref{fig:combined_image}d highlights the five most common space group combinations found among polymorphic pairs.
The pair (71, 225) appears most frequently, suggesting a structural correspondence between these groups.
This trend supports the notion that polymorphism tends to arise within predictable symmetry groupings tied to underlying atomic arrangements.
\par
Finally, Figure~\ref{fig:combined_image}e shows a clear trend where materials with lower average formation energy tend to have a low variance, indicating consistent and robust thermodynamic stability across their configurations.
In contrast, materials with higher or positive average formation energy exhibit much greater variance, suggesting a wider spread in stability among their possible polymorphs.
Since positive formation energy generally indicates instability, the high variance in this region implies the presence of both highly unstable and potentially metastable structures.
This relationship can guide materials discovery by highlighting systems with a low mean and low variance as strong candidates for synthesis, while those with high variance may still offer metastable phases worth exploring.
Out of 3,025 polymorph formulas that have reported formation energy, 17.42\% have a variance of more than 0.01, and only 3.44\% of them have a variance of more than 0.1.

\subsection*{Topological Correspondence Across Polymorphic Space Group Pairs}
To understand the polymorph transitions, we identified the most frequent polymorph space group pairs in the Materials Project database. Among them, the top five are: (71, 225) in ternary materials, followed by (63, 221), (63, 139), (139, 221), and (221, 225) in binary materials, as illustrated in Figure~\ref{fig:combined_image}d.

\begin{figure}[H]
    \centering
    \begin{minipage}[t]{0.32\textwidth}
        \includegraphics[width=\textwidth]{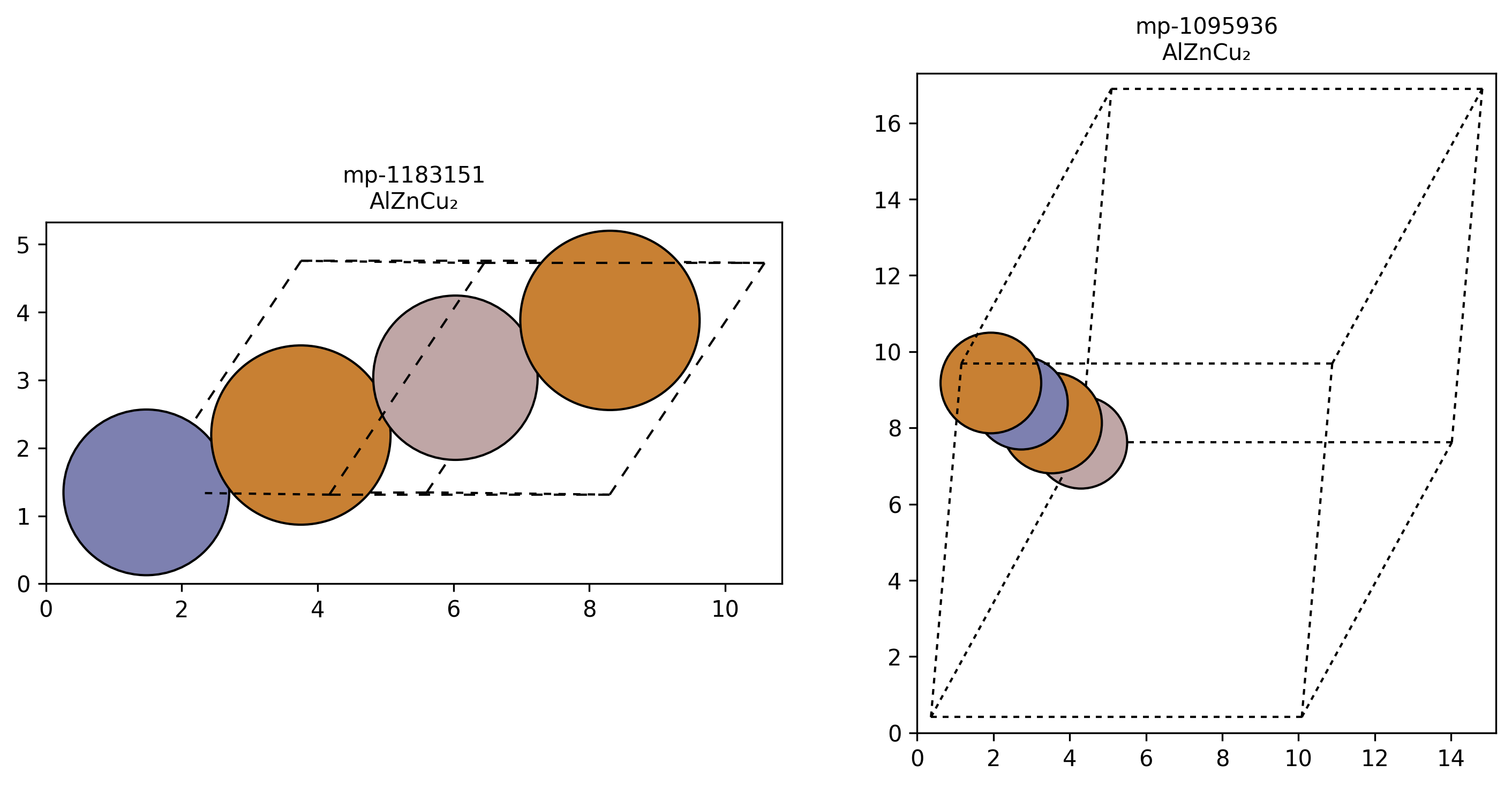}
        \caption*{(a)}
    \end{minipage}
    \hfill
    \begin{minipage}[t]{0.32\textwidth}
        \includegraphics[width=\textwidth]{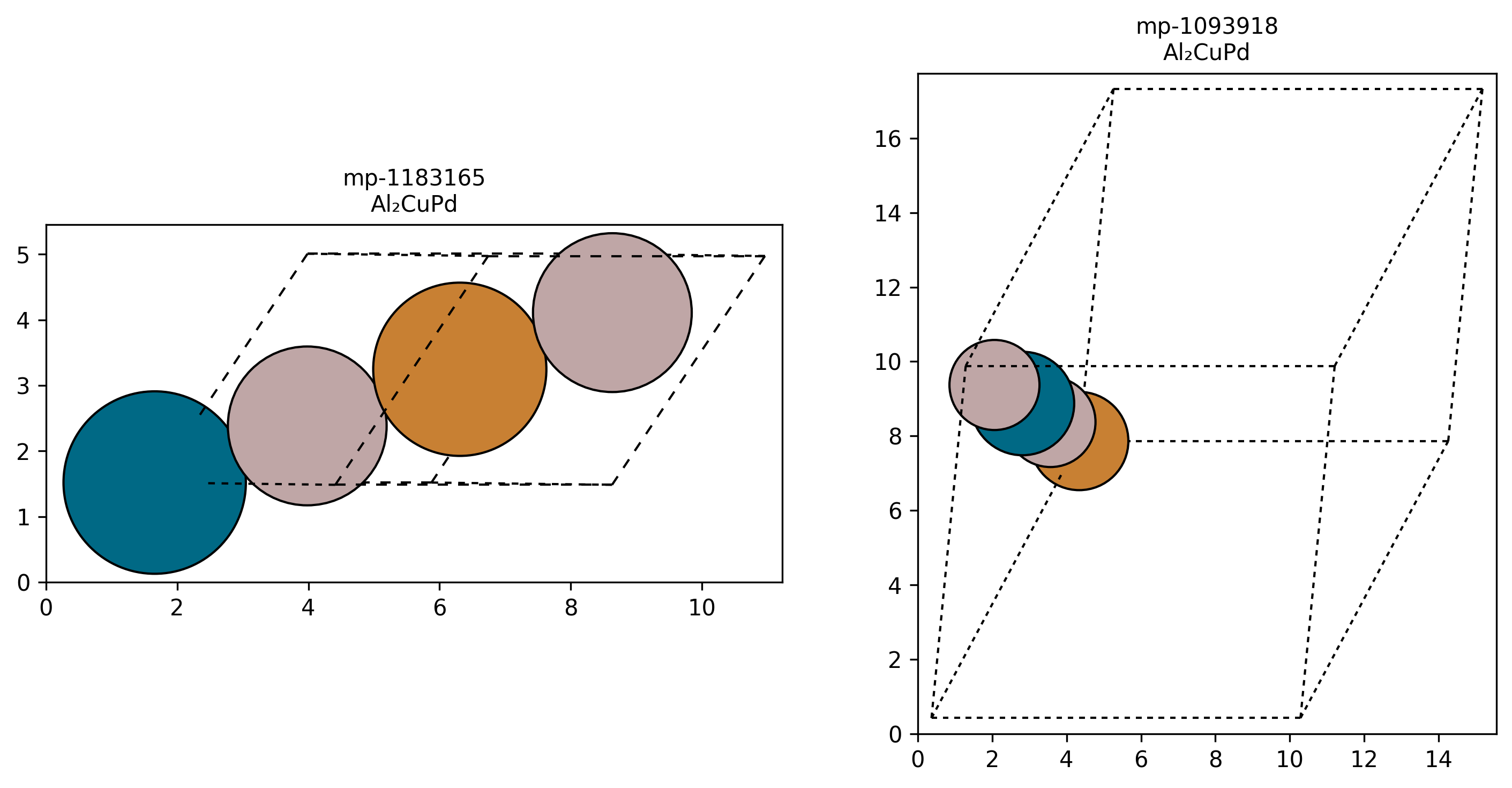}
        \caption*{(b)}
    \end{minipage}
    \hfill
    \begin{minipage}[t]{0.32\textwidth}
        \includegraphics[width=\textwidth]{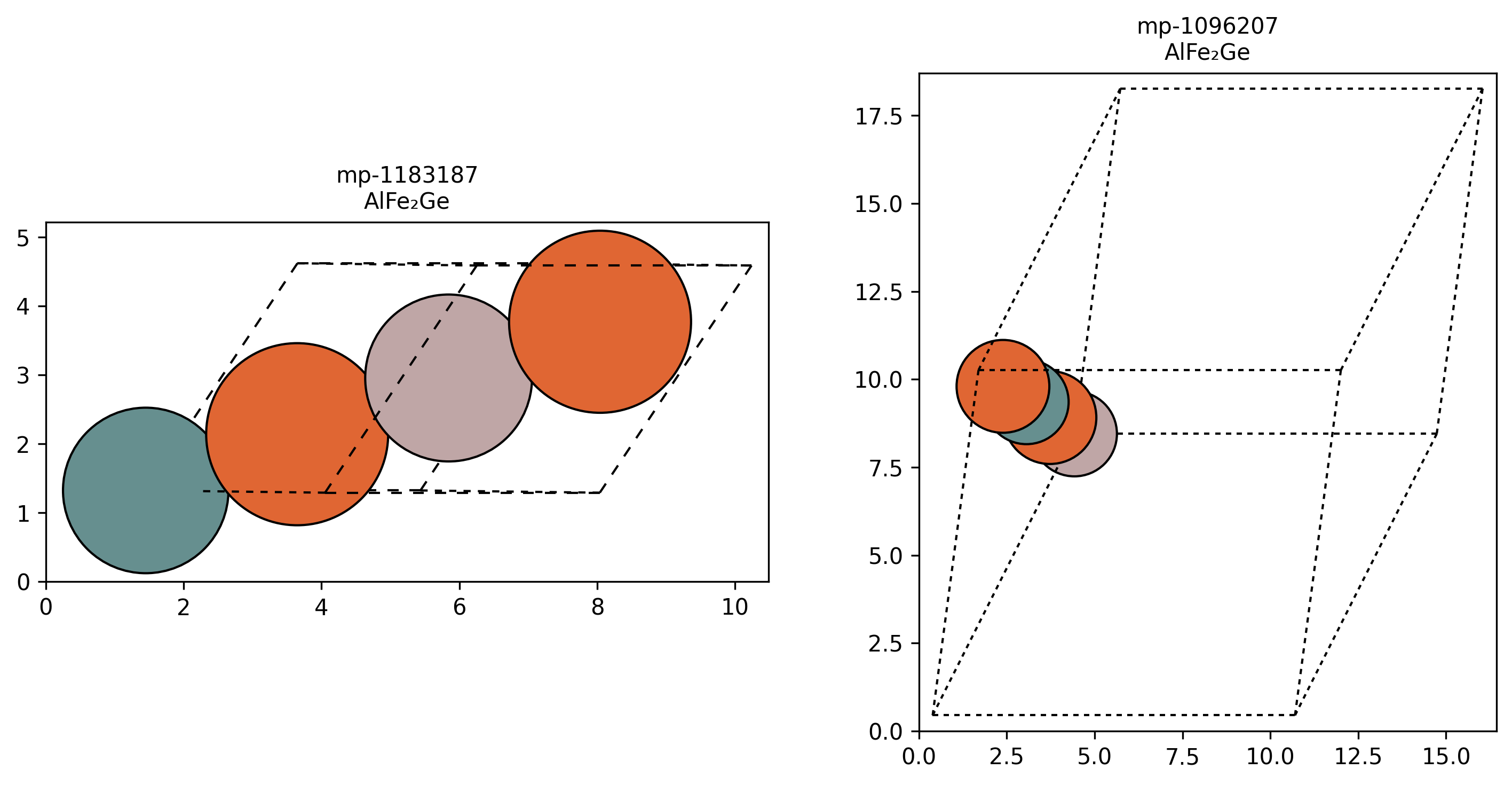}
        \caption*{(c)}
    \end{minipage}
    \caption{Each image shows two polymorphs of the same compound in different space groups (71,225). Now, (a), (b), and (c) all exhibit this consistency in pairwise pattern. While the polymorphs within a single compound are not identical, comparison across compounds reveals that all space group 225 structures share a similar topology, as do all space group 194 structures. This suggests that some space groups share recurring topological motifs across different compounds.}
    \label{fig:sgpair_similarity}
\end{figure}

A notable structural trend emerges when analyzing these frequent space group pairs (see Figure \ref{fig:sgpair_similarity}). For instance, consider the space group pair (71,225) as in the figure. Upon deeper investigation, we found that for many materials exhibiting polymorphism between these space groups, the atomic arrangements within the same space group (either 71 or 225) tend to be highly similar when their prototypes match. In addition to that, we found pair-wise similarity across different compounds. When filtering polymorphs structures with  space group pairs, we observe that when space group 225 structures share similar atomic configurations both locally and globally, their corresponding polymorphs in space group 71 also tend to show similarities. This suggests that polymorphism in these systems is often governed by symmetry-lowering distortions rather than fundamental changes in atomic arrangement. Similar patterns are observed in other common pairs, such as 63 and 221, 63 and 139, and 139 and 221. Although space groups like 71 and 225 frequently co-occur, this does not imply a direct phase transition between them. Our relaxation experiments using M3GNet \cite{chen2022universal} confirm the absence of spontaneous transitions across these pairs.

\FloatBarrier

\subsection{Polyhedral Topology and Local Structures}

\vspace{1mm}

\textbf{Polyhedra Construction }During polyhedron construction, the unit cell is sufficient for constructing polyhedra and capturing all unique coordination environments, as it contains at least one representative of each symmetrically distinct atomic site.
Therefore, analyzing the unit cell ensures that all polyhedron types present in the symmetrized structure are accounted for.
During the coordination number calculation for each site, the following atoms are ignored: O, S, N, F, Cl, Br, I, and H.
Thus, none of the constructed polyhedra have any of these atoms as a central atom.
To find the nearest neighbors of each site, we used CrystalNN provided by the Pymatgen package \cite{pan2021benchmarking}.
We considered the following types of polyhedra: Triangular, Tetrahedral, Trigonal bipyramidal, Octahedral, Pentagonal bipyramidal, Cubic/Square antiprismatic, Tricapped trigonal prismatic, Bicapped square antiprismatic, Pentagonal antiprismatic, and Icosahedral/Cuboctahedral.

\begin{figure}[h]
    \centering
    \begin{subfigure}[b]{0.32\textwidth}
        \centering
  
      \includegraphics[width=\textwidth, trim=20 20 20 20, clip]{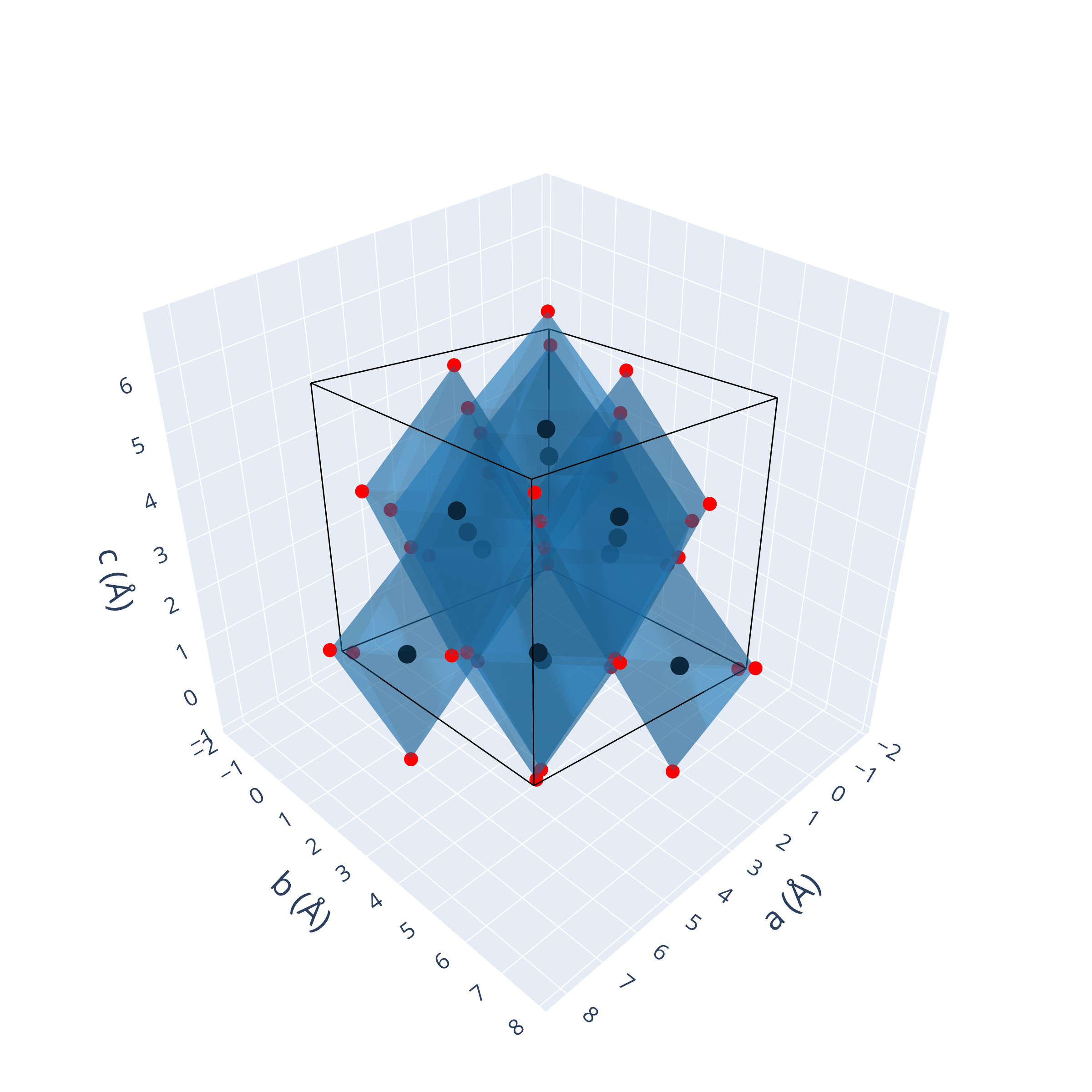}
    \end{subfigure}
    \hfill
    \begin{subfigure}[b]{0.32\textwidth}
        \centering
        \includegraphics[width=\textwidth, trim=20 20 20 20, clip]{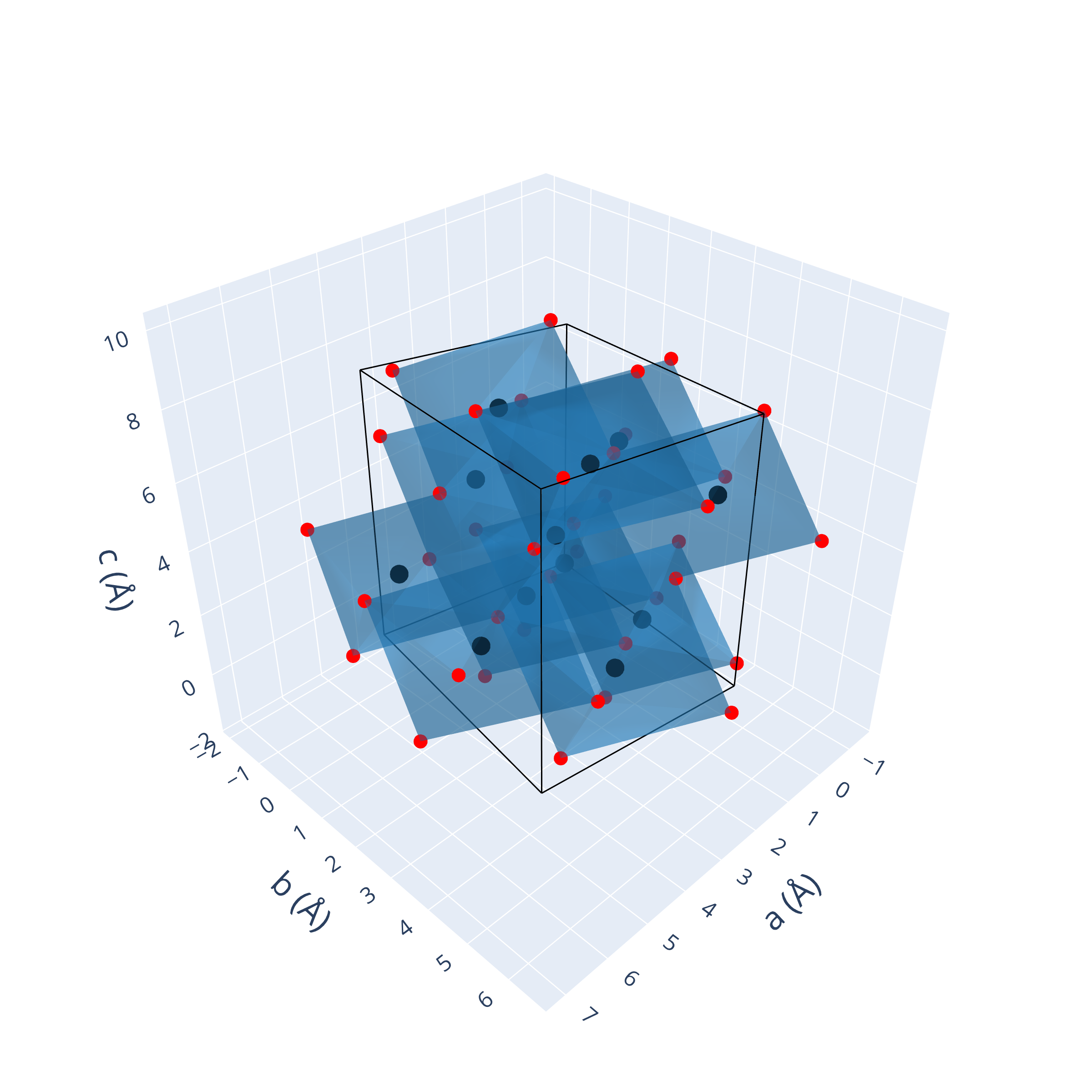}
    \end{subfigure}
    \hfill
    \begin{subfigure}[b]{0.32\textwidth}
        \centering
        \includegraphics[width=\textwidth, trim=20 20 20 20, clip]{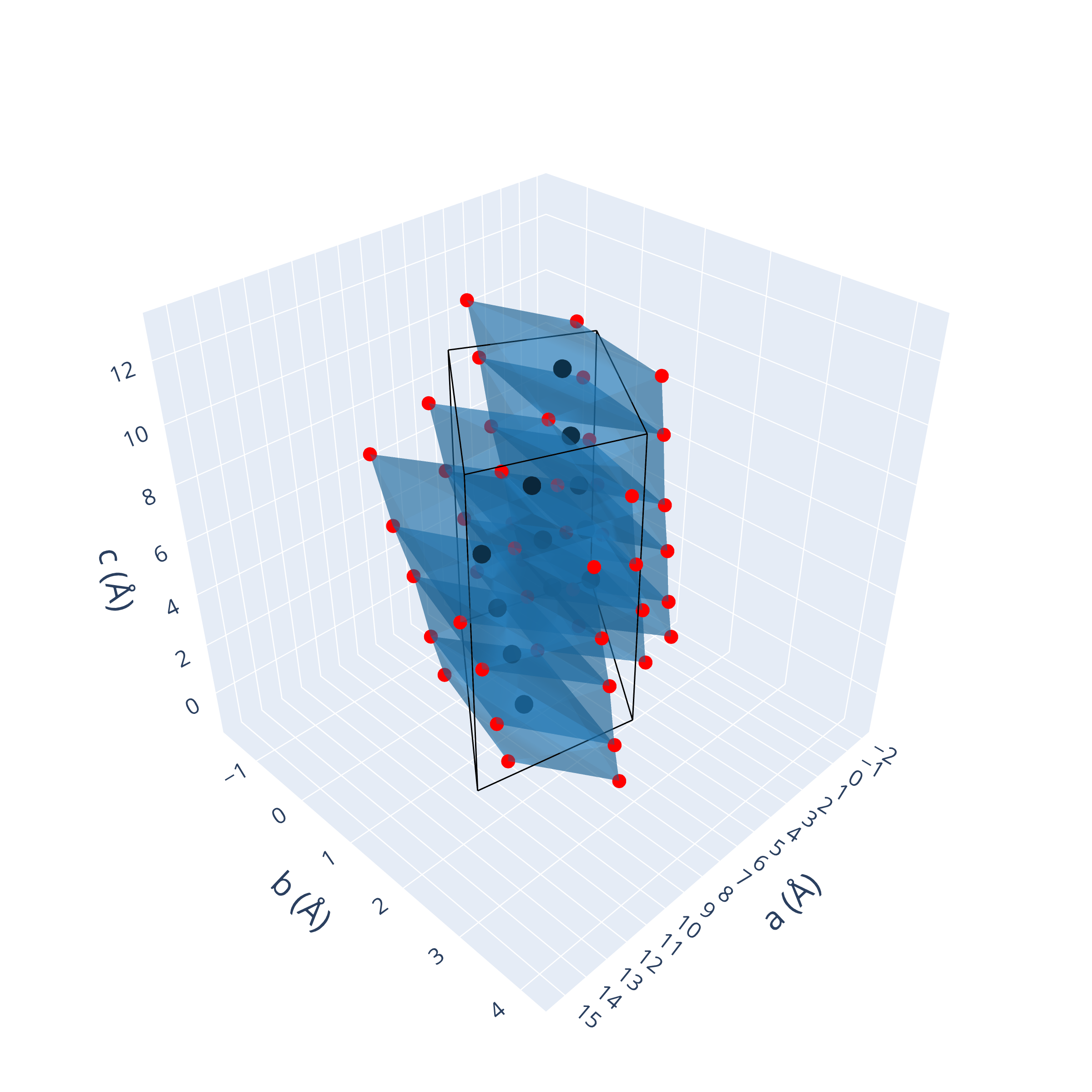}
    \end{subfigure}
    \caption{Visualization of polyhedra from 
randomly selected Li$_{7}$Mn$_{2}$(CoO$_{4}$)$_{3}$ structures. Despite variations in atomic arrangements, all structures share a common octahedral polyhedron. Non-site atoms are red; Site atoms (Li and Mn) are dark-colored.}
    \label{fig:polyhedra}
\end{figure}

\begin{figure}[H]
    \centering
    \begin{subfigure}[b]{0.48\textwidth}
        \centering
        \includegraphics[width=\textwidth]{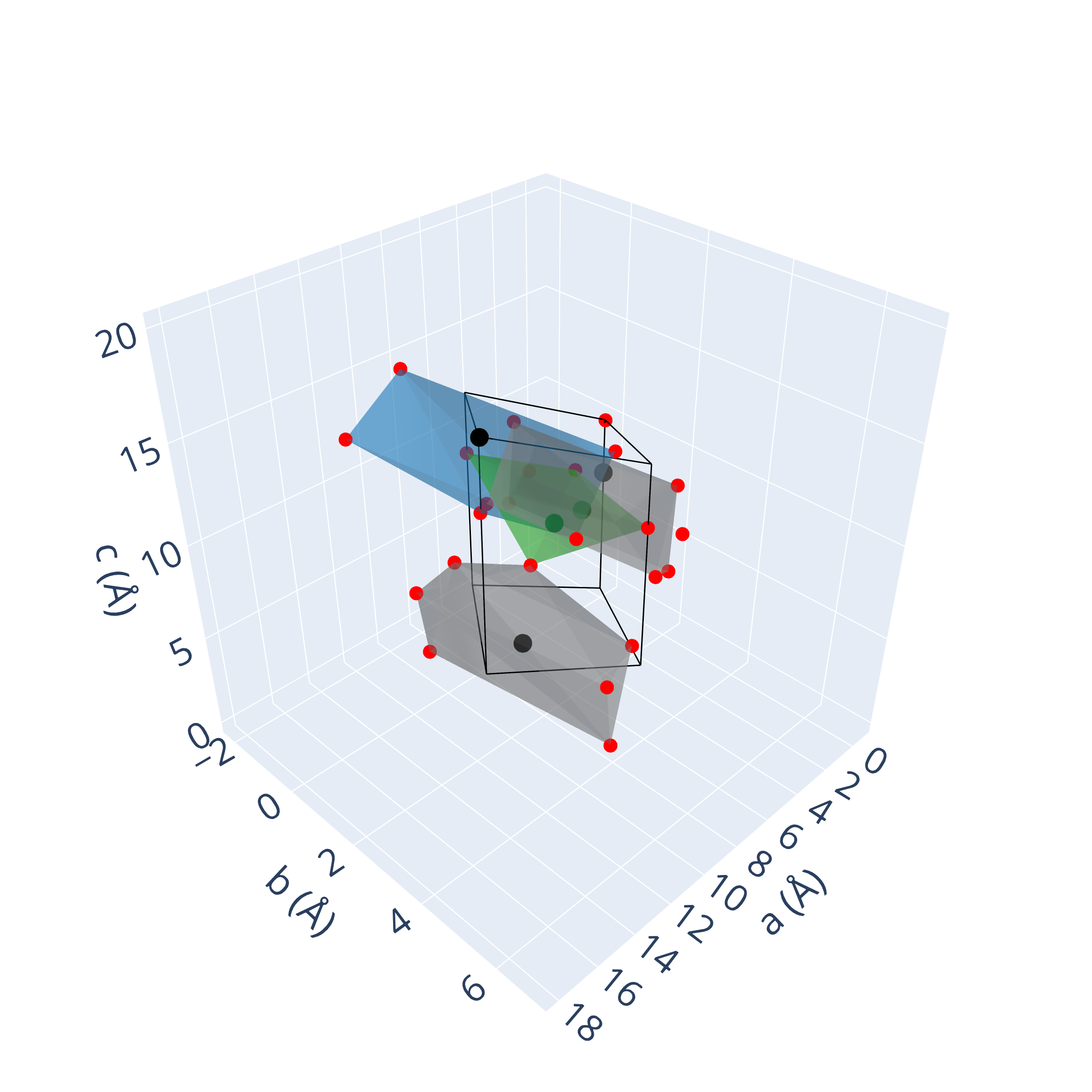}
        \caption{}
        \label{fig:subfig1}
    \end{subfigure}
    \hfill
    \begin{subfigure}[b]{0.48\textwidth}
        \centering
        \includegraphics[width=\textwidth]{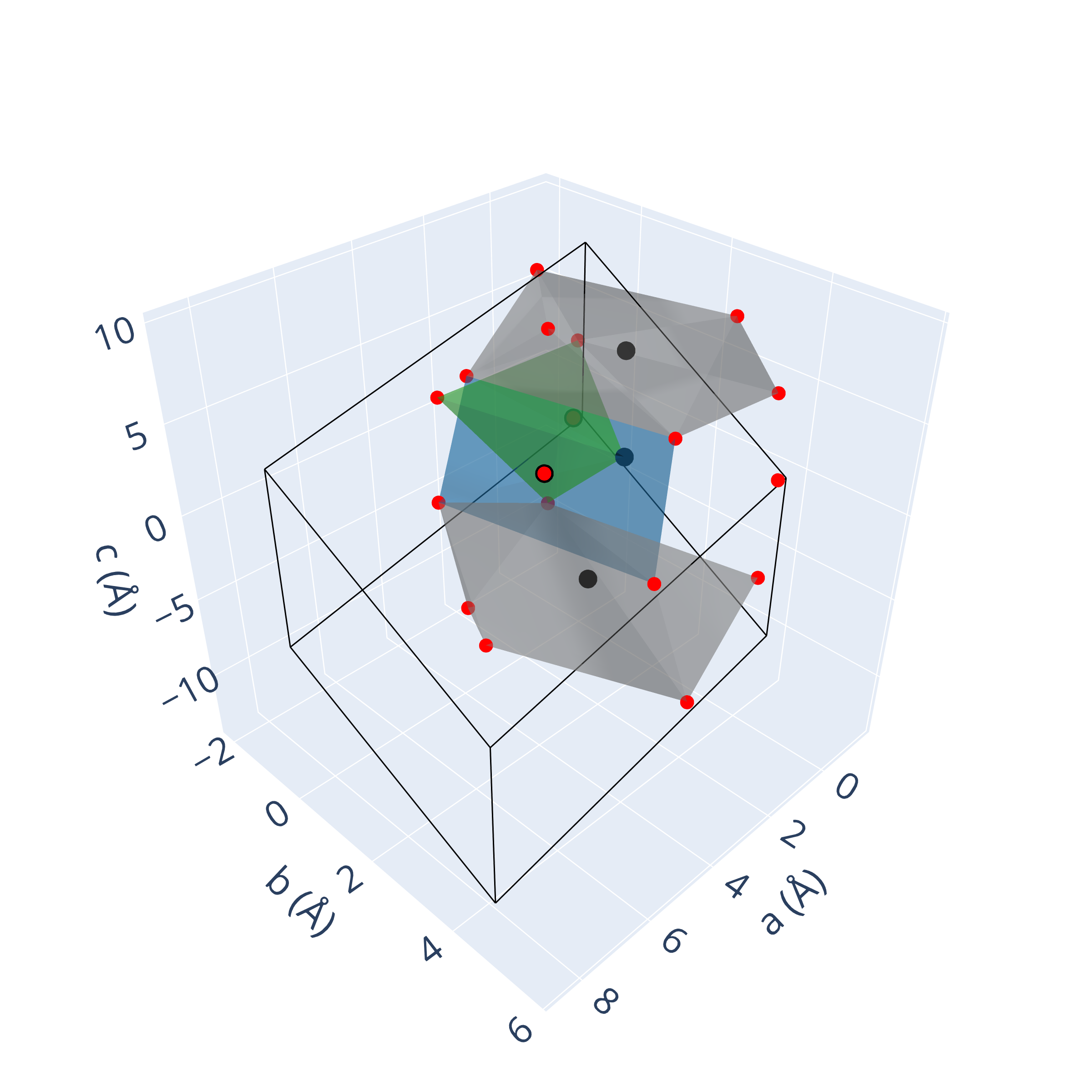}
        \caption{}
        \label{fig:same_polyhedra}
    \end{subfigure}
   
 \caption{Polyhedral representations of two different crystal structures of the formula Pb$_{4}$SeBr$_{6}$.
Both of the structures have an identical set of polyhedra. The Pb atoms have trigonal, tetrahedral, and pentagonal polyhedra, and Se has a triangular polyhedron.
This same set holds true for both structures.
}
    \label{fig:polyhedra-identical}
\end{figure}

Among the polymorphic materials analyzed, 3,247 formulas, which account for more than half of the entire polymorph dataset, were found to have the same number of polyhedra across their polymorphs.
Furthermore, 644 formulas exhibited not only the same number but also the same shape of polyhedra (e.g., octahedral, tetrahedral) across their corresponding polymorphs. While some formulas have only one type of polyhedron (see Figure \ref{fig:polyhedra} for Li$_{7}$Mn$_{2}$(CoO$_{4}$)$_{3}$),
it is also seen that some polymorph formulas have an identical set of polyhedra for all their structures.
For example, Pb$_{4}$SeBr$_{6}$ has trigonal, pentagonal, tetragonal, and triangular polyhedra in two of its polymorphs (space groups are different: 8 and 44), as can be seen in Figure \ref{fig:polyhedra-identical}.

This highlights how shared local environments can span diverse global symmetries, yet remain structurally coherent.
We report 180 polymorph formulas that have an identical set of polyhedron types (more than 3 types) within their corresponding polymorphs in supplementary file section 3.

\subsection{Polyhedral Connectivity in Li–Mn–Co–O Systems}
\begin{figure}[h]
    \centering
    \begin{subfigure}[b]{0.48\textwidth}
        \centering
        \includegraphics[width=\linewidth]{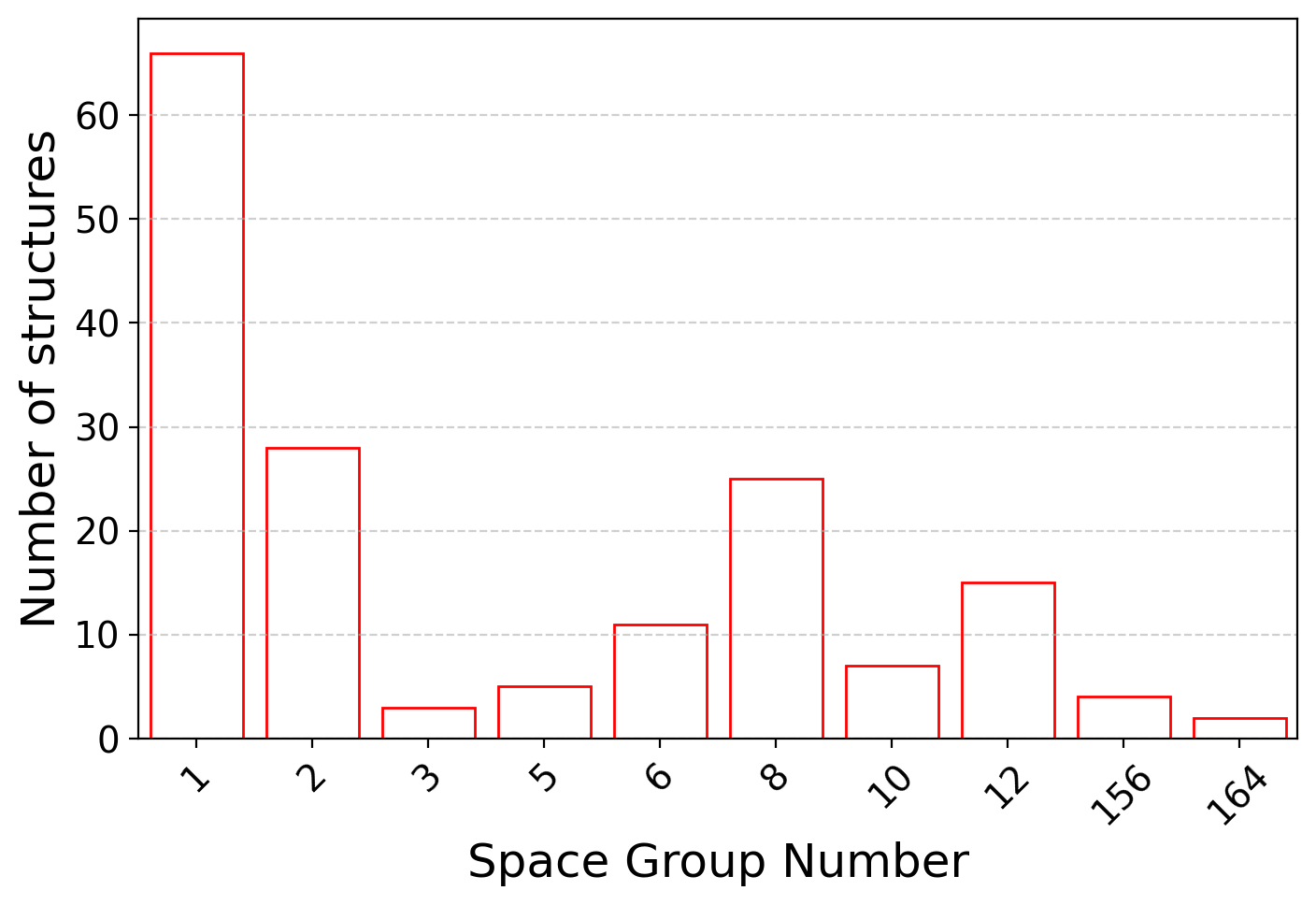}
        \caption{Space group distribution for Li$_{7}$Mn$_{2}$(CoO$_{4}$)$_{3}$.}
        \label{fig:sg_num_Li7Mn2}
    \end{subfigure}
    \hfill
    \begin{subfigure}[b]{0.48\textwidth}
        \centering
        \includegraphics[width=\linewidth]{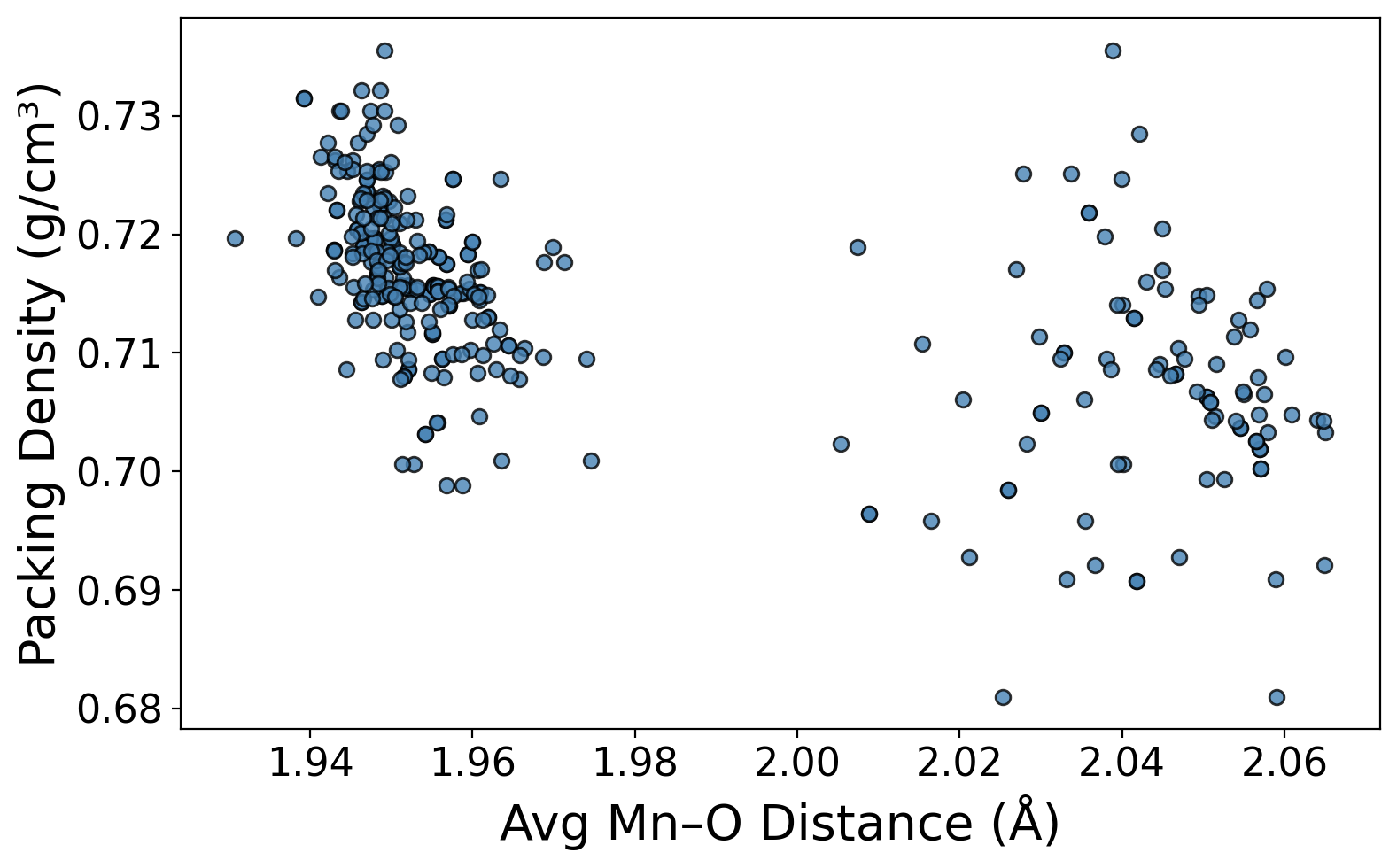}
        \caption{Relationship between average Mn–O bond distance and packing density across polymorphs.}
        \label{fig:mno_distance_density}
  
  \end{subfigure}
    \caption{Structural and geometric trends in the polymorphs of Li$_{7}$Mn$_{2}$(CoO$_{4}$)$_{3}$.}
    \label{fig:li7mn2_structural_trends}
\end{figure}

We examined Li$_{7}$Mn$_{2}$(CoO$_{4}$)$_{3}$, which has the largest number of reported polymorphic structures in our dataset.
Out of its 166 reported structures, 165 exhibit an identical octahedral polyhedron (see Figure \ref{fig:polyhedra}), with only one exception containing a bipyramidal geometry.
A more detailed analysis of Li$_{7}$Mn$_{2}$(CoO$_{4}$)$_{3}$ reveals that these preserved octahedra are not geometrically identical as space groups of these structures vary (see Figure \ref{fig:sg_num_Li7Mn2}) as well as the Mn–O bond lengths separate into two distinct families (see Figure \ref{fig:mno_distance_density}), one with shorter, more compact octahedra and another with longer, more relaxed ones.
These geometric variations strongly correlate with packing density: polymorphs with shorter Mn–O bonds tend to exhibit higher packing density, while those with longer bonds are more loosely packed.
This demonstrates that even when the polyhedron type is preserved, subtle geometric distortions at the local level can significantly influence the global structure.
This same trend is observed in other transition-metal oxides such as  Li$_{7}$Mn$_{5}$CoO$_{12}$, Li$_{4}$Mn$_{3}$CoO$_{8}$, and Li$_{3}$MnCoO$_{5}$.
In all systems, Mn-centered octahedra are consistently present across polymorphs, yet they exhibit a similar bimodal distribution in bond lengths (see Figure \ref{fig:mno_distance_density}).
In each case, shorter Mn–O bonds correspond to denser polymorphs.
While shorter Mn–O bonds generally correlate with higher packing density, this relationship is not strictly monotonic.
In several polymorphs, long-bond octahedra are also found in densely packed structures, suggesting that global packing efficiency arises not only from local polyhedral compression but also from the overall structural arrangement and contributions of other atomic species.
Additionally, among the 17 unique formulas containing Li, Mn, Co, and O, 14 were found to contain only a single type of polyhedron across all their polymorphs.
This further underscores the role of local structural motifs in constraining polymorphic diversity.
We performed a detailed polyhedral motif analysis for a set of selected Li-Mn-Co oxide formulas.
For each formula, we evaluated the percentage of structures exhibiting only octahedral coordination and those exhibiting a single type of polyhedron.
The summary is provided in Table~\ref{tab:li_mn_co_polyhedral}.

\begin{table}[h!]
\centering
\caption{Polyhedral motif statistics for selected Li-Mn-Co oxide formulas. The results indicate that Li–Mn–Co oxides overwhelmingly favor purely octahedral coordination, with most compositions forming single-polyhedron frameworks, highlighting their strong local structural uniformity across polymorphs.}
\label{tab:li_mn_co_polyhedral}
\begin{tabular}{lcccc}
\hline
\textbf{Formula} & \textbf{Total Materials} & \textbf{Only Octahedral (\%)} & \textbf{Single Polyhedron (\%)} \\ \hline
Li$_3$MnCoO$_5$ & 47 & 100.00 & 100.00 \\
Li$_4$Mn$_3$CoO$_8$ & 26 & 96.15 & 96.15 \\
Li$_5$Mn$_2$CoO$_8$ & 18 & 100.00 & 100.00 \\
Li$_7$Mn$_2$(CoO$_4$)$_3$ & 166 & 99.40 & 99.40 \\
Li$_7$Mn$_4$CoO$_{12}$ & 71 & 88.73 & 88.73 \\ \hline
\end{tabular}
\end{table}

Having polymorphs does not indicate a narrower packing density. The SiO$_{2}$ that has five different types of polyhedra have a wider packing density, while Li$_{7}$Mn$_{2}$(CoO$_{4}$)$_{3}$, having only one type of polyhedron in 165 out of 166 structures shows a very narrow packing density distribution. Although it has only one type of polyhedron, it has different space groups, most of which tend towards lower symmetry constraints, as can be seen in Figure \ref{fig:sg_num_Li7Mn2}.

\subsection{Prototype-wise Polyhedral Trends}
Building upon this detailed analysis of polyhedral motifs in Li–Mn–Co–O systems, we next explored how such local coordination trends manifest across broader prototype families.
To understand the local coordination environments in different material classes, we performed a polyhedral motif analysis across various prototype groups.
For each material, we determined whether (i) all detected polyhedra were of a single shape and (ii) whether all detected polyhedra were exclusively octahedral.
The results are summarized in Table~\ref{tab:polyhedral_statistics}.

\begin{table}[h!]
\small
\centering
\caption{Polyhedral motif statistics across different prototype groups in polymorph structures. While quaternary and ternary systems show the highest fraction of structures containing only octahedra (34.30\% and 29.25\%, respectively), the Rocksalt/Zincblende/CsCl group exhibits the largest proportion of single-polyhedron networks (79.39\%), suggesting strong motif uniformity. In contrast, other prototypes show minimal octahedral or single-polyhedron dominance, reflecting greater local environment diversity.}
\label{tab:polyhedral_statistics}
\begin{tabular}{lrrrr}
\hline
\textbf{Prototype} & \textbf{Total Materials} & \textbf{Only Octahedral (\%)} & \textbf{Single Polyhedron (\%)} \\ \hline
Binary & 3277 & 9.61 & 54.38 \\
Fluorite/Rutile & 1896 & 22.47 & 52.11 \\
Perovskite & 1293 & 14.93 & 19.49 \\
Quaternary & 2767 & 34.30 & 40.33 \\
Rocksalt/Zincblende/CsCl & 1504 & 25.00 & 79.39 \\
Ternary & 7931 & 29.25 & 48.81 \\
Other & 381 & 1.57 & 2.62 \\ \hline
\end{tabular}
\end{table}

\subsection*{Polymorphism in Material Prototypes}
We then analyzed polymorphism from the perspective of different material types, such as binary materials, ternary materials, different building blocks, and different topological views in the context of space groups.
Out of 7,799 binary structures, we found that 2,044 of them have polymorphs, and for ternary structures, 3,320 out of 25,866 structures show polymorphism.
Unlike the space group pair analysis, where we could find structural similarities based on prototypes, we could not find any patterns because the structures of binary materials constitute diverse types of atomic orientations as well as polyhedron types.
Categorizing the binary structures into their well-known prototypes such as Rocksalt $AB$, Fluorite $AB_{2}$, and $AB_{3}$ compounds also did not reveal structural similarities in terms of polyhedron types, coordination numbers, or polyhedron numbers.
Similarly, for ternary materials, categorizing the structures into well-known prototypes such as Perovskite ($ABX_3$), Spinel ($AB_2X_4$), Chalcopyrite ($ABX_2$), Pyrochlore ($A_2B_2X_7$), Scheelite ($ABX_4$), and Olivine ($A_2BX_4$) was unhelpful in retrieving similar structures. Given the limitations of prototype-based classification in revealing polymorphic trends, we turned to a topology-driven novel approach to uncover structural relationships beyond symmetry constraints.

\subsection{Topological Mapping of Polymorphic Materials}
 To study the polymorphism distribution of crystals, we clustered them to find different types of polymorph structures by mapping them in a topological polymorph space.
This topology is not constrained by symmetry.

\textbf{Step 1: Topology-based Graph} \\
This is achieved using Algorithm \ref{alg:polyhedron_graph}.

\begin{algorithm}
\caption{Constructing Polyhedron Connectivity Graphs }
\label{alg:polyhedron_graph}
\begin{algorithmic}[1]
\REQUIRE A set of polyhedral blocks extracted from a crystal structure
\ENSURE A graph $G = (V, E)$ representing connections between polyhedra

\STATE Initialize an empty graph $G$ with nodes $V \leftarrow \{\text{polyhedra}\}$

\FORALL{pairs of polyhedra $(P_i, P_j)$ such that $i \ne j$}
    \IF{$P_i$ and $P_j$ share exactly 4 vertices}
        \STATE Add edge $(P_i, P_j)$ to $E$ with label \texttt{face-quad}
    \ELSIF{$P_i$ and $P_j$ share exactly 3 vertices}
        \STATE Add edge $(P_i, P_j)$ to $E$ with label \texttt{face-tri}
   
 \ELSIF{$P_i$ and $P_j$ share exactly 2 vertices}
        \STATE Add edge $(P_i, P_j)$ to $E$ with label \texttt{edge}
    \ELSIF{$P_i$ and $P_j$ share exactly 1 vertex}
        \STATE Add edge $(P_i, P_j)$ to $E$ with label \texttt{point}
    \ENDIF
\ENDFOR

\RETURN $G$
\end{algorithmic}
\end{algorithm}

\textbf{Step 2: Graph Embedding.} \\
Each topology-based graph derived from Step 1 is converted into a fixed-length feature vector by computing degree histograms for each connection type: edge-sharing (2 shared atoms), triangular face-sharing (3 atoms), 
and square face-sharing (4 atoms). 
For each type, a histogram counts how many polyhedra have 0, 1, 2, ..., $n$ such connections.
These histograms are normalized and concatenated to form a single vector representing the polyhedral connectivity pattern of the structure. \\

\textbf{Step 3: Dimensionality Reduction via t-SNE \cite{maaten2008visualizing}.}
To visualize the similarity between structures based on their polyhedral topology, the embedding vectors are projected into two dimensions using t-distributed stochastic neighbor embedding (t-SNE).
In this 2D space, structures with similar local polyhedron connection graphs appear close together, enabling unsupervised grouping and visual identification of topological clusters.

\FloatBarrier

\subsubsection{Test Cases on Perovskites}

\begin{figure}[th!]
\centering

\begin{subfigure}[t]{0.3\textwidth}
    \centering
    \includegraphics[width=\linewidth]{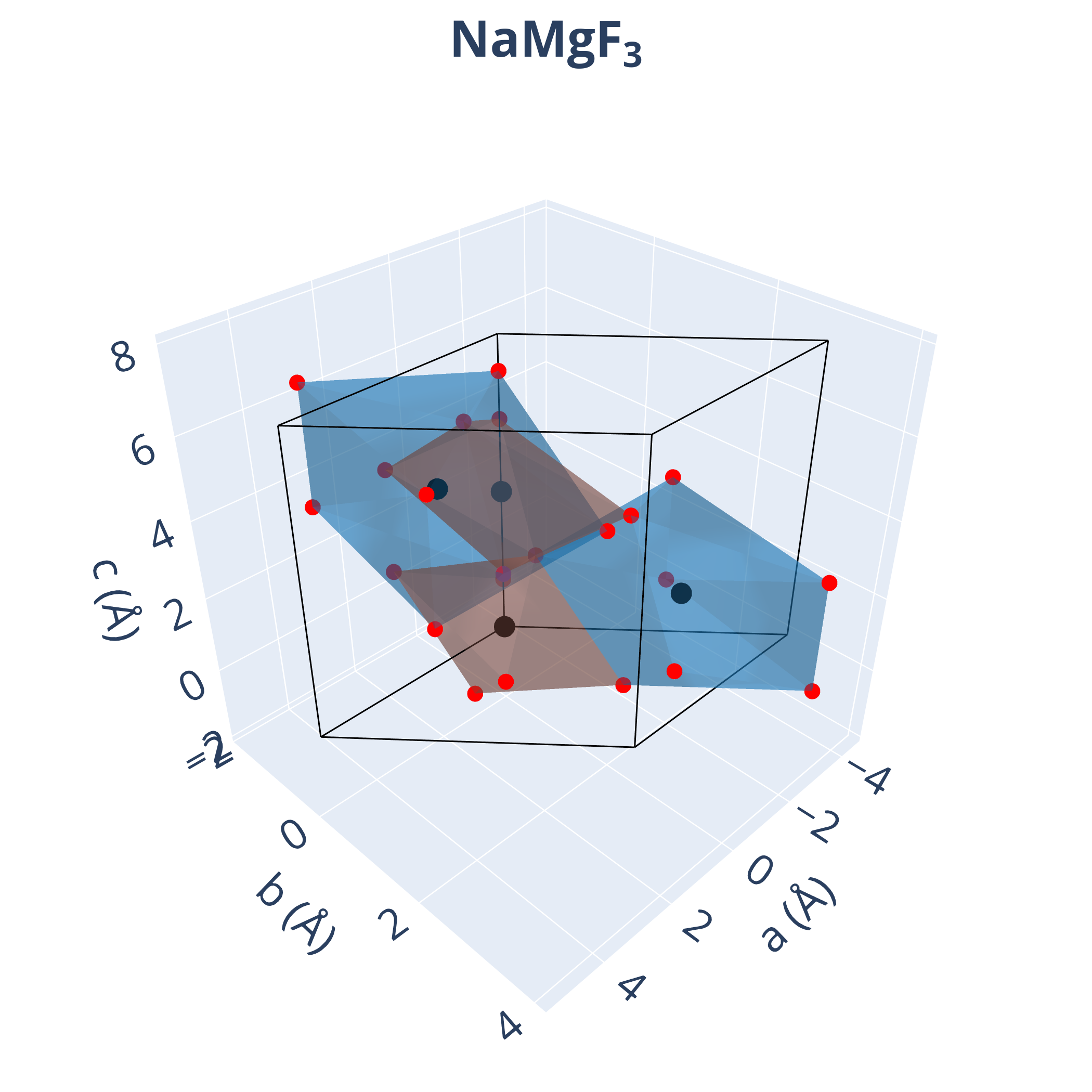}
    \caption{NaMgF$_3$ (mp-12948)}
    
\end{subfigure}
\hfill
\begin{subfigure}[t]{0.3\textwidth}
    \centering
    \includegraphics[width=\linewidth]{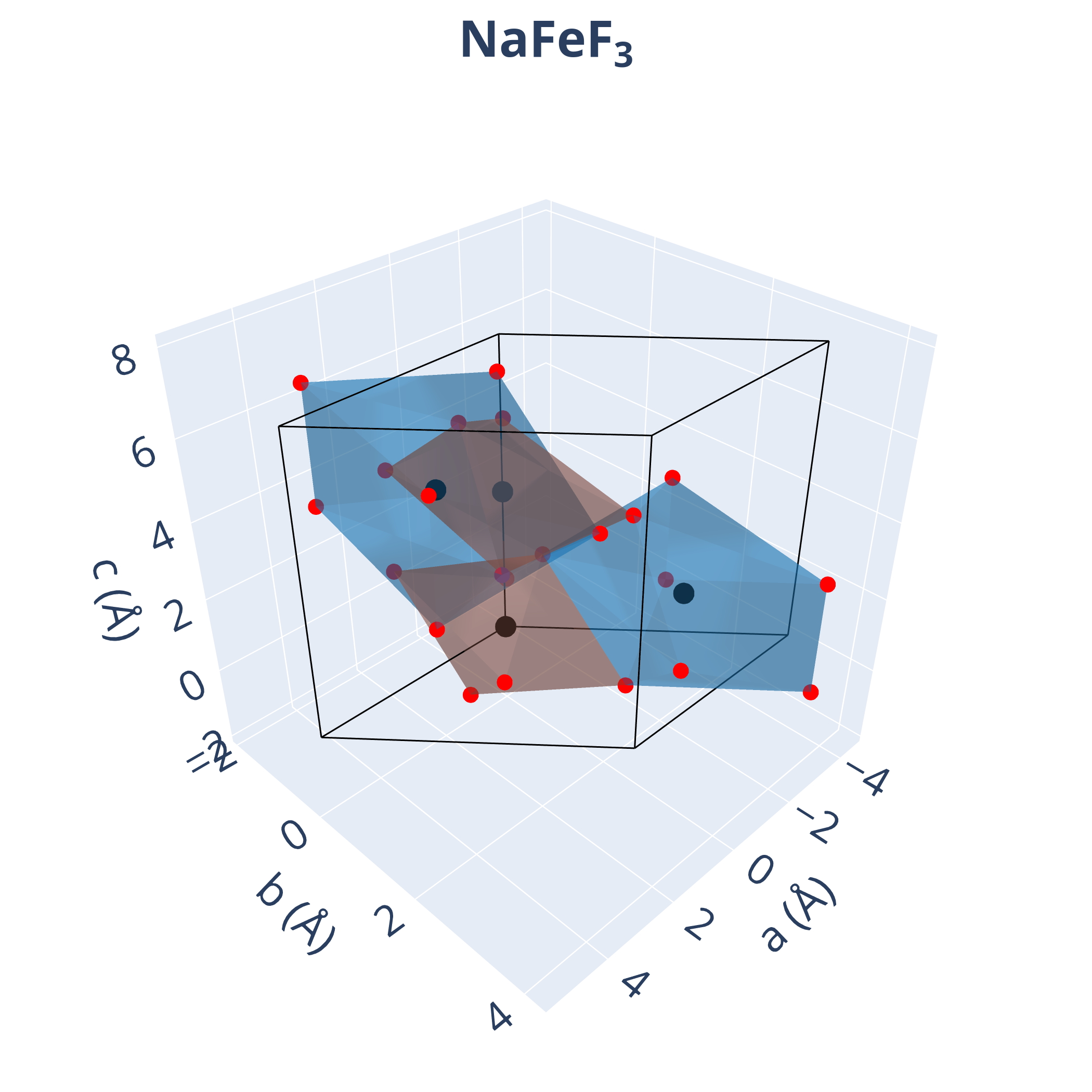}
    \caption{NaFeF$_3$ (mp-1078217)}
\end{subfigure}
\hfill
\begin{subfigure}[t]{0.3\textwidth}
    \centering
    \includegraphics[width=\linewidth]{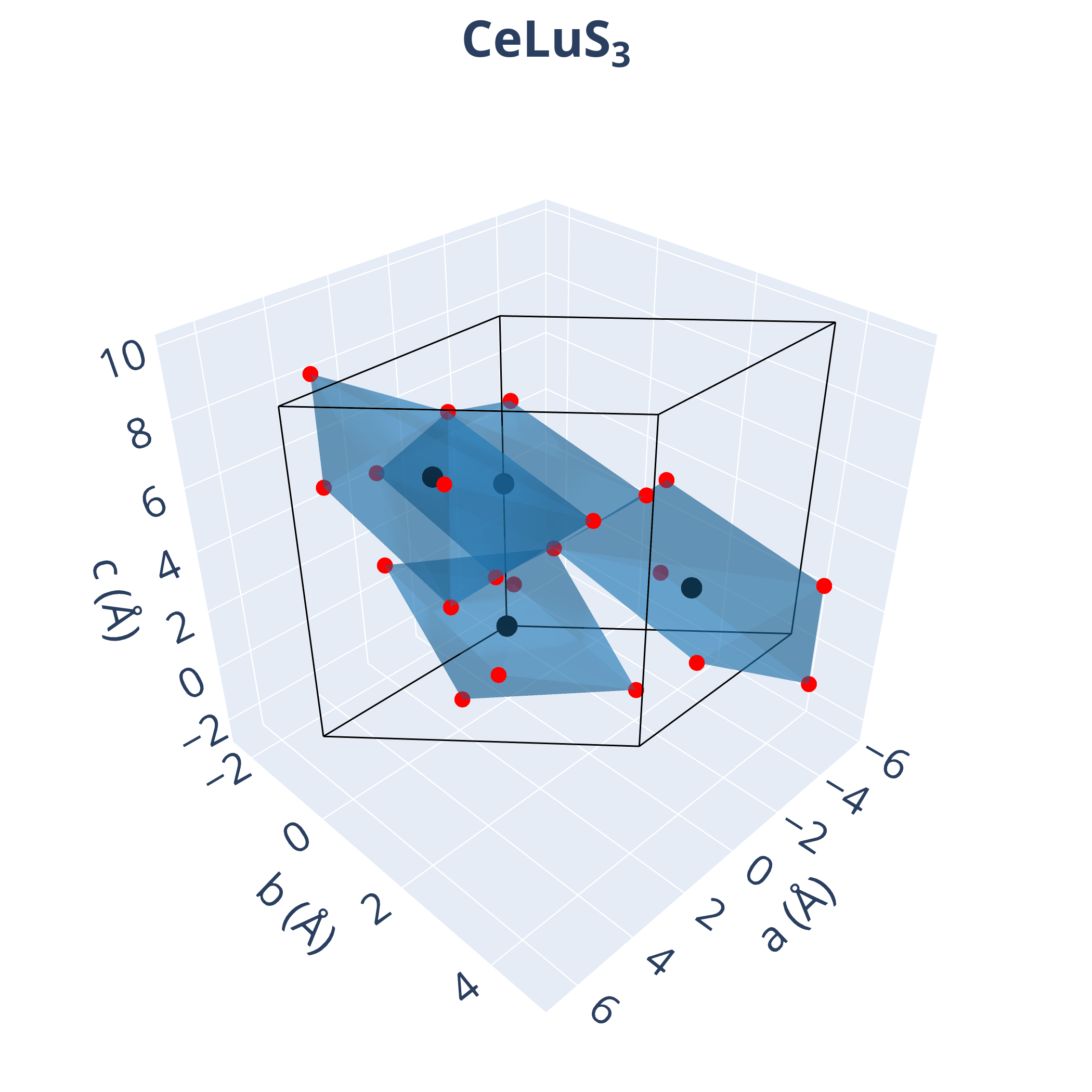}
    \caption{CeLuS$_3$ (mp-1205797)}
\end{subfigure}

\vspace{0.5cm}

\begin{subfigure}[t]{0.3\textwidth}
    \centering
    \includegraphics[width=\linewidth]{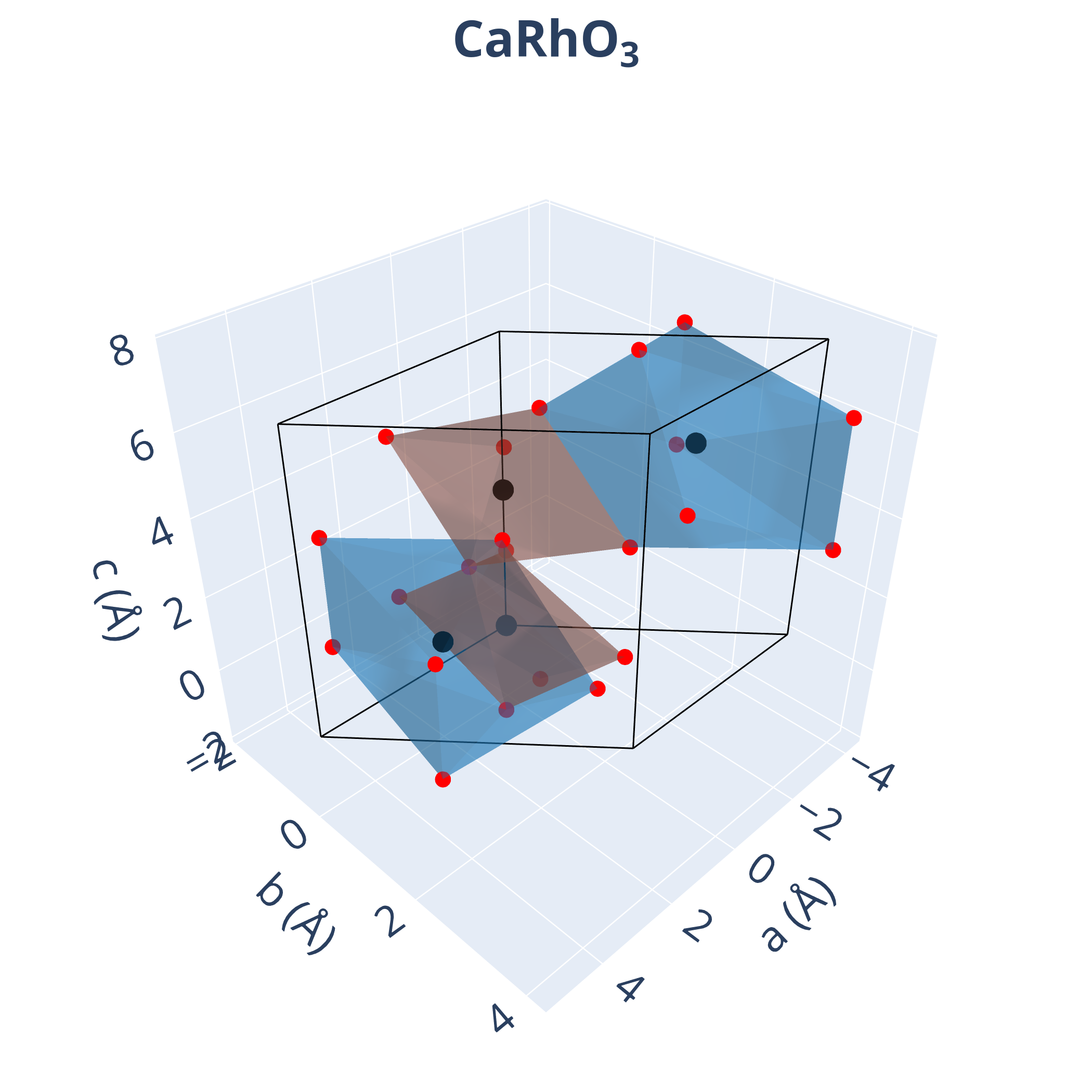}
    \caption{CaRhO$_3$ (mp-1078659)}
\end{subfigure}
\hfill
\begin{subfigure}[t]{0.3\textwidth}
    \centering
    \includegraphics[width=\linewidth]{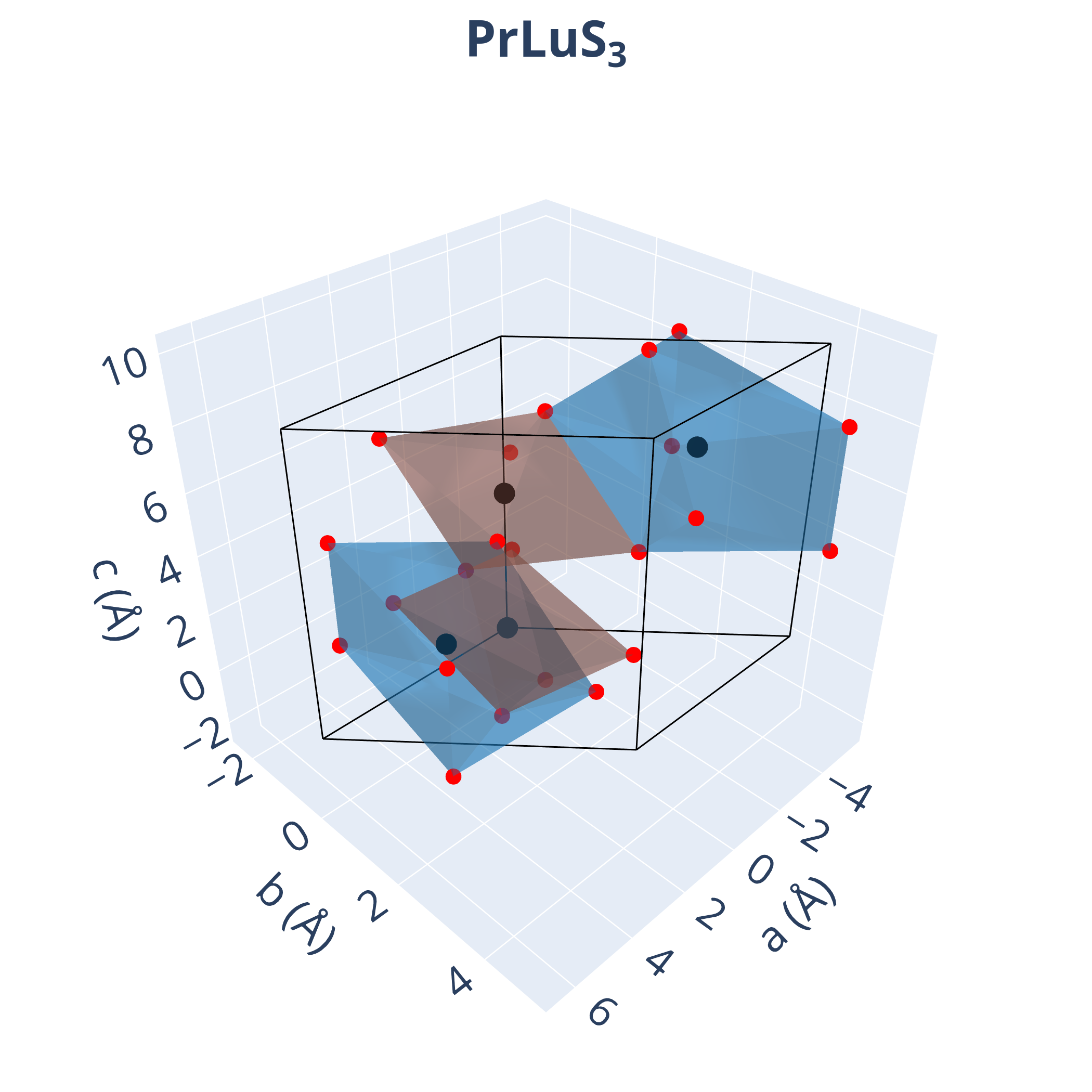}
    \caption{PrLuS$_3$ (mp-1078537)}
\end{subfigure}
\hfill
\begin{subfigure}[t]{0.3\textwidth}
    \centering
    \includegraphics[width=\linewidth]{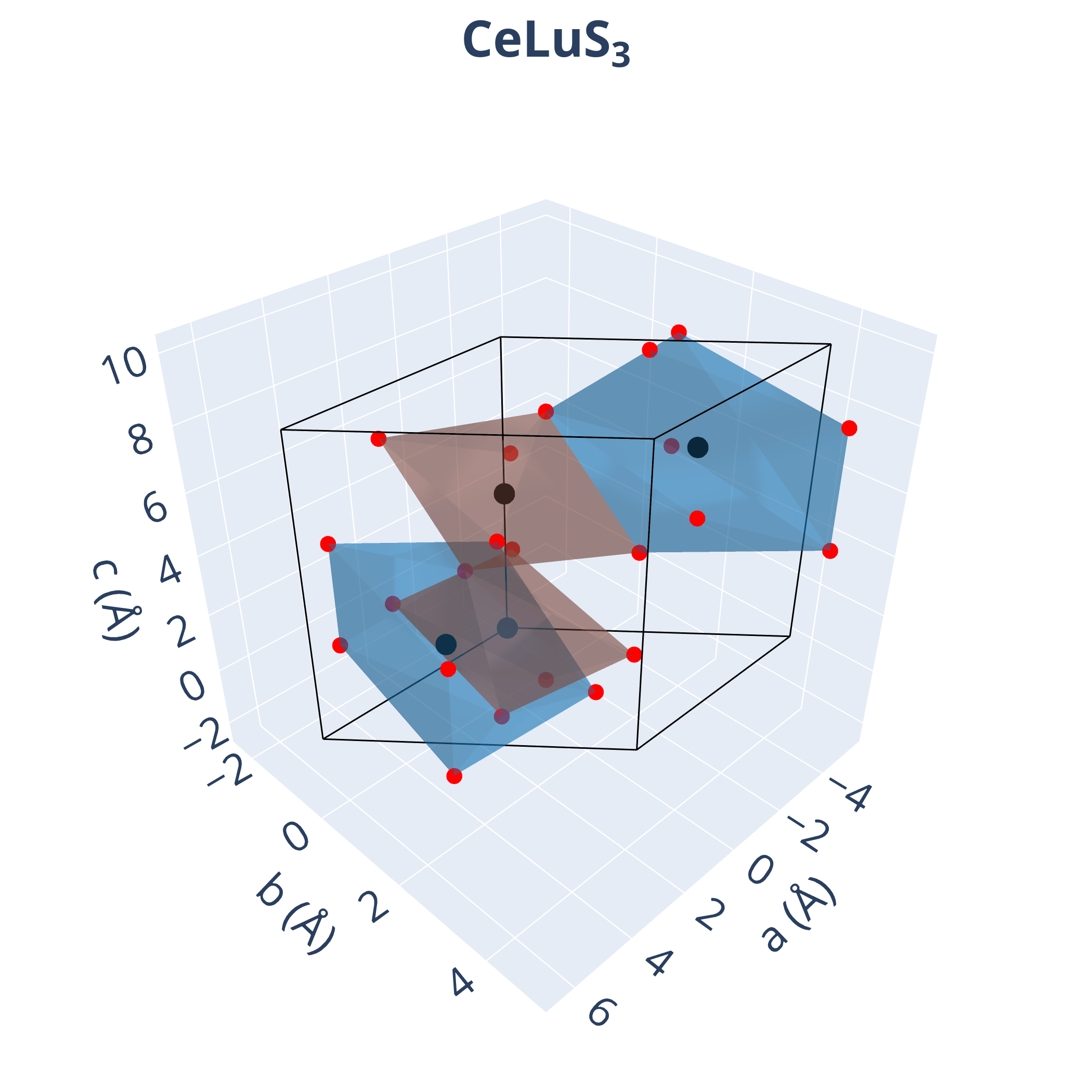}
    \caption{CeLuS$_3$ (mp-1092225)}
\end{subfigure}

\caption{Polyhedral environments for NaMgF$_3$ and nearby structures in the t-SNE connectivity space.
Each structure is labeled with its chemical formula and Materials Project ID.
Although the topologies of these structures do not look exactly alike, they have very similar trends in terms of connectivity types.
These structures contain both octahedral and cubic/square antiprismatic polyhedra. In all six cases, four polyhedra form a repeating face-sharing sequence between alternating types: a cubic/square antiprismatic unit shares a face with an octahedron, which in turn shares a face with another cubic/square antiprismatic polyhedron, and so on, creating a linked pattern of alternating connectivity.}
\label{fig:tsne_namgf3_neighbors}
\end{figure}

\textbf{NaMgF$_3$} is a well-explored perovskite that has been studied for decades to understand its phase transition~\cite{martin2006phase, zhao1993thermal}.
We found 2 of its polymorphs in the database with space group numbers 63 and 221. Plotting them reveals that these two are located in separate regions, while the plot also enables us to find very similar yet distinct perovskite structures, as can be seen in Figure \ref{fig:tsne_namgf3_neighbors}.\\

\textbf{NbAgO$_{3}$} has 4 polymorphs reported in the database with space groups 63, 127, 221, and 65. When plotting the structure with space group 63, the figure shows that it is clustered around similar structures, all of which have very similar topology (i.e., polyhedron distribution).
Thus, it paves the way to learn about similar structures (see supplementary Figure S3).
Furthermore, these similar structures are not constrained by symmetry only, as structures with the same space group can be located in different clusters, and structures with different space groups can stay within the same cluster.

\subsubsection{Test Cases on Binary (Fluorite/Rutile) Materials}
In this test case, we focus on the binary compound $CaCd_{2}$, which exhibits polymorphism within the Fluorite/Rutile structural family (see  Figure supplementary S4). Figure S4 presents two polymorphs of $CaCd_{2}$ (mp-1444 and mp-1078), each followed by its closest neighbors in the t-SNE space derived from polyhedral connectivity graphs.
Notably, the retrieved neighbors span diverse compositions and space groups. However, they all share remarkably similar local coordination topologies.
This includes consistent polyhedral geometries and recurrent edge- and face-sharing patterns.
These structural analogs, which are not evident from symmetry labels alone, highlight the strength of the proposed graph representation in capturing transferable topological features.

\subsection{Test Cases on Quaternary polymorphs} In supplementary Figure S5, we present a set of quaternary compounds clustered closely in the t-SNE space, centered around \textbf{Te$_2$Mo$_3$(SeS)$_2$ (mp-1026002)}.
Each structure exhibits repeating layers of distorted triangular polyhedron stacked along the $c$-axis, forming columnar arrangements.
Notably, \textbf{Mo$_2$W(Se$_2$S)$_2$}, \textbf{MoW$_2$(Se$_2$S)$_2$}, and \textbf{Te$_2$Mo$_3$(SeS)$_2$} (both polymorphs) maintain strong topological alignment with the reference structure, evident from the visual similarity in the number, orientation, and layering of polyhedra.
\textbf{Te$_4$Mo(WS)$_2$} shows a deviation in polyhedral connectivity but still retains key topological motifs, while \textbf{Te$_4$Mo$_2$WS$_2$} bridges structural features from both Te-rich and Mo-rich analogs.
These observations reinforce that polyhedral topology can serve as a more sensitive descriptor than symmetry alone, capturing continuity across chemically diverse quaternary phases.

\subsection{Oxidation state analysis of polymorphs}
\begin{figure}[htbp]
    \centering
    \includegraphics[width=0.9\textwidth]{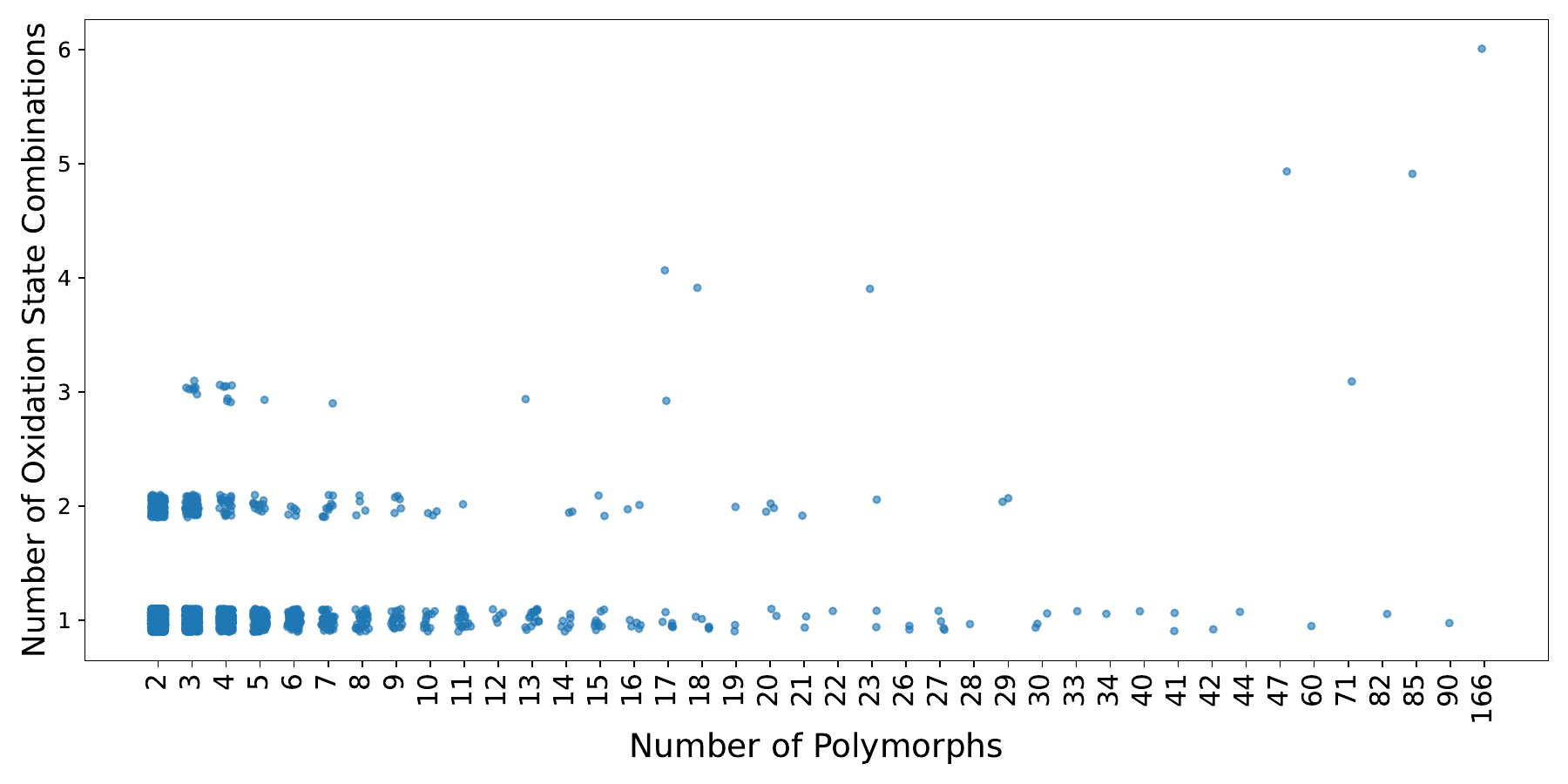}

    \caption{Scatter plot showing the number of oxidation state combinations observed for each polymorph formula (with more than one distinct crystal structure) in the dataset.
Each point represents a formula, where the x-axis indicates the number of its polymorphs, and the y-axis shows the number of distinct oxidation state combinations observed across those polymorphs.}
    \label{fig:os_distribution}
\end{figure}

To understand how the formation of polymorphic structures depends on different oxidation states, we analyzed oxidation state consistency across polymorph structures from the Materials Project database using its API.
For each formula with more than one distinct structure, we counted the number of oxidation state configurations and visualized the results in the scatter plot shown in Figure \ref{fig:os_distribution}. 
Our findings show that the majority of formulas exhibit a single oxidation state combination regardless of polymorphic diversity, suggesting structural transitions are not commonly accompanied by redox changes. However, a subset of materials (Li$_{7}$Mn$_{2}$(CoO$_{4}$)$_{3}$) with high polymorphic counts exhibit substantial variation in oxidation states, indicating potential redox flexibility. These materials warrant further study for applications in catalysis or electrochemical storage.
Despite having multiple distinct structures, most formulas (3,725 formulas) exhibit only a single oxidation state configuration among their polymorphs.
Only a small fraction of formulas (337 samples) have two oxidation state configurations, only 19 formula contains three oxidation state configurations and just 6 of them contain 4 or more configurations among its polymorphs. This demonstrates that polymorphs of materials do not have to have distinct oxidation state configurations. They can have different polyhedrons and their varying arrangements, while their polyhedron motifs can have the same oxidation state configurations.

\section{Discussion}

This work applies big data topological analysis to the polymorphic crystal structures to understand their inherent patterns, motifs, and graphic representation for global mapping. By analyzing the most frequent space group pairs observed among polymorphic materials, we found a consistent trend: 
for a given space group pair (e.g., 71 and 225), structures belonging to the same space group often exhibit highly similar atomic arrangements when they share the same prototype.
This structural consistency was evident across multiple frequent polymorph pairs.
However, relaxation studies using M3GNet showed that these structures do not spontaneously transition between the paired space groups (e.g., from 71 to 225 or vice versa), suggesting that the observed polymorphism is not a result of a direct phase transformation.
Instead, it points toward an underlying structural constraint, where specific atomic motifs are statistically more likely to reappear across certain space groups, even in the absence of a dynamic pathway connecting them.
This emphasizes the role of prototype-driven structural similarity in governing polymorphic relationships.

Our analysis reveals that despite variations in global symmetry across polymorphs, local structural motif, particularly the number and shape of polyhedra are often remarkably conserved.
This structural persistence suggests that polymorphism is governed not only by thermodynamic factors but also by deep geometric constraints rooted in coordination chemistry.
Such findings underscore the importance of local atomic environments in shaping polymorphic landscapes and offer a new perspective for predicting and designing polymorphs based on polyhedral consistency.
In addition to these findings, we proposed a polyhedron connectivity graph-based topology-aware embedding strategy to generate a global map of polymorphic materials.
This embedding consistently clustered together polymorphs and topologically similar structures, even when their space groups differ.
Such clustering confirms that polyhedral connectivity captures meaningful structural similarity that transcends symmetry classification.
This observation supports the idea that topological descriptors can serve as robust indicators of polymorphic behavior, especially when symmetry-based comparisons fall short.

We further conducted subgroup-wise mapping by analyzing representative structural families, including perovskite, spinel, chalcopyrite, pyrochlore, scheelite, and olivine.
Across these families, we observed that polymorphs often preserve consistent local polyhedral arrangements despite global differences in symmetry or atomic packing.
These trends reinforce that polymorphism is shaped not only by energetic considerations but also by deep geometric constraints embedded in coordination environments. Complementing the structural analysis, we examined the oxidation state diversity across polymorphic formulas. Interestingly, even as the number of reported polymorphs increases for a given chemical formula, the diversity of oxidation state combinations remains low for most cases. Structural polymorphs typically preserve the same electronic configuration (oxidation states), implying that redox-driven phase transitions are relatively rare in the dataset analyzed. This reinforces the view that many polymorphic transformations occur without significant changes in valence states, highlighting the structural—rather than electronic nature of most polymorphic variability observed.
Together, our polymorph topological analysis results demonstrate that topology-guided graph-based representations offer a powerful framework for identifying and comparing polymorphs.
By emphasizing structural motifs and polyhedral connectivity over strict symmetry, our approach provides a new pathway for understanding and predicting polymorphic landscapes across diverse material classes.

\section{Data Availability}

The polymorph crystal material id are available in the github repository (see section Code Availability, dataset folder and mp-id column name) which are downloaded from the Materials Project Database at \url{http://www.materialsproject.org}. 

\section{Code Availability}

The analysis source code can be found at \url{http://www.github.com/usccolumbia/polymorphism}

\section{Contribution}
Conceptualization, J.H.; methodology, S.D., J.H., N.M., S.S.O., R.D., N.F. C.W.; software, S.D., N.F.; resources, J.H.; writing—original draft preparation, S.D., N.M., S.S.O., R.D., N.F., C.W.,J.H.;
writing—review and editing, J.H; visualization, S.D., N.F. and J.H.; supervision, J.H.; funding acquisition, J.H.
\section*{Acknowledgement}
The research reported in this work was supported in part by the National Science Foundation under grants 2311202,2110033 and 2320292. The views, perspectives, and content do not necessarily represent the official views of the NSF.
\bibliographystyle{unsrt}  
\bibliography{references}

\end{document}